\documentclass[sigconf,10pt]{acmart}

\usepackage[a-1b]{pdfx}
\usepackage{lmodern}
\usepackage{listings}
\usepackage{color}
\usepackage{epsfig,makecell}                                      
\hypersetup{citecolor=blue,linkcolor=blue}                        
\usepackage{thmtools}
\usepackage{thmbox}
\usepackage[toc,page]{appendix}
\usepackage{endnotes,microtype,xspace,fancyvrb,multirow} 
\usepackage{epigraph} 

\usepackage{booktabs}      
\usepackage{tabularx}
\usepackage{array,underscore,relsize}                             
\usepackage[T1]{fontenc}                                          
\usepackage{times}                                                
\usepackage{enumitem}                                             
\usepackage[labelfont=bf,font=normal,skip=1pt]{caption}           
\usepackage[export]{adjustbox}                                    
\usepackage{breakurl}                                             
\usepackage{pifont}                                               
\usepackage{tablefootnote}                                        
\usepackage{bbding}                                        
\usepackage[linesnumbered,vlined,boxed,commentsnumbered]{algorithm2e}
\DontPrintSemicolon
\newcommand{\ra}[1]{\renewcommand{\arraystretch}{#1}}
\usepackage[normalem]{ulem}
\usepackage[hang,flushmargin]{footmisc}
\usepackage{textcomp}
\usepackage{graphicx}

\usepackage[scaled]{beramono}
\usepackage{subfig}
\usepackage{float}
\usepackage{listings}
\usepackage{courier}
\usepackage{color}
\usepackage{colortbl}
\usepackage{stfloats}

\usepackage{stfloats}

\usepackage[compact]{titlesec}
\titleformat*{\section}{\Large\bfseries}
\titleformat*{\subsection}{\normalsize\bfseries}
\titleformat*{\subsubsection}{\normalsize\bfseries}
\titlespacing{\section}{0pt}{3ex}{1.5ex}
\titlespacing{\subsection}{0pt}{2ex}{1ex}
\titlelabel{\thetitle\hspace{-1mm}\quad}

\lstdefinelanguage{CUDA}{
    keywords={__global__, __device__, __shared__, __host__, __constant__, __restrict__},
    keywordstyle=\color{blue}\bfseries,
    ndkeywords={int, float, double, char, unsigned, void},
    ndkeywordstyle=\color{red}\bfseries,
    identifierstyle=\color{black},
    sensitive=false,
    comment=[l]{//},
    morecomment=[s]{/*}{*/},
    commentstyle=\color{purple}\ttfamily,
    stringstyle=\color{red}\ttfamily,
    morestring=[b]",
}

\hypersetup{
  colorlinks,
  linkcolor={red!50!black},
  citecolor={blue!50!black},
  urlcolor={blue!80!black}
}

\newcommand{\fig}[1]{Figure{~\ref{#1}}}

\newcommand{\one}{\texttt{\uppercase\expandafter{\romannumeral1}}}
\newcommand{\two}{\texttt{\uppercase\expandafter{\romannumeral2}}}

\newcommand{\stitle}[1]{\vspace{1.2ex}\noindent{\bf #1}\,}
\newcommand{\nospacestitle}[1]{\noindent{\bf #1}\,}

\newcommand{\syslong}{{\textsc{PhoenixOS}}}
\newcommand{\sys}{\textsc{{P}hOS}}

\newcommand{\sysshort}{{\textsc{PhOS}}}

\definecolor{list_gray}{rgb}{0.5,0.5,0.5}
\definecolor{list_blue}{rgb}{0.13,0.13,1}
\definecolor{list_green}{rgb}{0,0.5,0}
\definecolor{list_mauve}{rgb}{0.58,0,0.82}
\usepackage{tikz}

\usepackage{balance}
\copyrightyear{2025}
\acmYear{2025}
\setcopyright{acmlicensed}\acmConference[SOSP '25]{ACM SIGOPS 31st Symposium on Operating Systems Principles}{October 13--16, 2025}{Seoul, Republic of Korea}
\acmBooktitle{ACM SIGOPS 31st Symposium on Operating Systems Principles (SOSP '25), October 13--16, 2025, Seoul, Republic of Korea}
\acmDOI{10.1145/3731569.3764813}
\acmISBN{979-8-4007-1870-0/2025/10}

\begin{document}


\title{\huge {\syslong}: Concurrent OS-level GPU Checkpoint and Restore \\with Validated Speculation}

\author{\Large
  \vspace{-1mm}
  Xingda Wei$^{\dagger\,1}$ \ \ Zhuobin Huang$^{\dagger\,2}$ \ \
  Tianle Sun$^1$ \ \ Yingyi Hao$^1$ \ \ Rong Chen$^{\ddagger\,1}$ \\
  Mingcong Han$^1$ \ \ Jinyu Gu$^1$ \ \ Haibo Chen$^1$ \\ \vspace{2mm}
  {{$^{1}$\,Institute of Parallel and Distributed Systems, Shanghai Jiao Tong University} \\ 
  {$^{2}$\,National University of Singapore}}
  \vspace{2mm}
}

\renewcommand{\shortauthors}{X. Wei, Z. Huang, T. Sun, Y. Hao, R. Chen, M. Han, J. Gu, and H. Chen}

\date{}


\begin{abstract}
\noindent
{\syslong} ({\sys}) is the first OS service that can concurrently checkpoint and restore (C/R)
GPU processes---a fundamental capability for critical tasks 
such as fault tolerance, process migration, and fast startup.
While concurrent C/R is well-established on CPUs, it poses unique challenges on GPUs 
due to their lack of essential features 
for efficiently tracing concurrent memory reads and writes,
such as specific hardware capabilities (e.g., dirty bits) 
and OS-mediated data paths (e.g., copy-on-write).

To ensure correct concurrent C/R, {\sys} proactively detects GPU memory reads and writes
through a two-step process: first, it speculates about GPU memory accesses 
based on the arguments used when launching GPU kernels; 
then, it validates these accesses efficiently at runtime using binary instrumentation.
With this validated speculation, {\sys} retrofits CPU-based concurrent C/R for GPUs
through software-based approaches, including soft copy-on-write, soft recopy, and soft on-demand restore.
{\sys} further proposes several GPU-aware techniques for efficient GPU C/R,
including coordinated checkpoint data transfer and execution context pool. 
For downstream tasks that use 
C/R for tolerating failures, migrating processes between machines, 
and accelerating cold starts in serverless computing,
{\sys} achieves orders of magnitude higher performance than state-of-the-art OS-level GPU C/R systems
like NVIDIA cuda-checkpoint. 

\end{abstract}


\keywords{GPU checkpoint and restore, concurrent checkpoint and restore, validated speculation\vspace{1mm}}

\maketitle
\pagestyle{empty}

\def\thefootnote{$\dagger$}\footnotetext{Xingda Wei and Zhuobin Huang contributed equally.
Part of the work was done when Zhuobin Huang was an intern at Institute of Parallel and Distributed Systems, Shanghai Jiao Tong University.}
\def\thefootnote{$\ddagger$}\footnotetext{Rong Chen is the corresponding author.}

\renewcommand{\thefootnote}{\arabic{footnote}}

\section{Introduction}
\label{sec:intro}

\nospacestitle{What is concurrent OS-level checkpoint and restore (C/R) and why it matters. \,}
OS-level checkpoint snapshots the execution of a running process
as an image, 
which the OS can later use to restore the process.
This is a foundational OS primitive with many key applications: 
cluster management systems rely on it to migrate tenant jobs~\cite{DBLP:journals/corr/abs-2202-07848,DBLP:conf/osdi/XiaoBRSKHPPZZYZ18,cuda-migration}
by first checkpointing the image to the target machine and then restoring from it. 
Serverless systems leverage OS-level restore to launch new processes
quickly~\cite{serverless-gpu-ali,aws-snapstart,DBLP:conf/asplos/DuYXZYQWC20}.
Cloud providers implement periodic checkpointing to provide fault tolerance 
against failures~\cite{DBLP:journals/corr/abs-2202-07848,DBLP:conf/sosp/Wu0M023}.

A key feature of OS-level checkpoint and restore is
that it is \emph{transparent} to processes, 
making it irreplaceable for both {functionality} and {efficiency}. 
For {functionality}, cloud providers need migrate tenant jobs to improve cluster utilization and 
other purposes~\cite{DBLP:journals/corr/abs-2202-07848,DBLP:conf/osdi/XiaoBRSKHPPZZYZ18,cuda-migration}.
Since these jobs operate as black boxes to the providers~\cite{cuda-migration}, 
migration can only be performed at the OS level. 
For {efficiency}, serverless platforms need to launch GPU processes quickly~\cite{ali-serverless-gpu}.
These processes contain PyTorch runtime states (e.g., compiler cache) that 
are tightly intertwined with OS information.
Rebuilding these states from scratch takes seconds; thankfully, OS-level C/R can 
significantly reduce this overhead to milliseconds~\cite{DBLP:conf/asplos/DuYXZYQWC20,DBLP:conf/osdi/WeiLW0Y0023,
DBLP:conf/asplos/Ustiugov0KBG21}.

Additionally,
OS-level C/R can also provide superior {usability}~\cite{DBLP:journals/corr/abs-2202-07848,microsoft-forge},
which is why major cloud providers (e.g., Microsoft Forge~\cite{microsoft-forge,DBLP:conf/eurosys/GuptaKKVGKRS24})
employ it for fault tolerance.
The key rationale is that performing efficient and correct fault tolerance
through user-level C/R is challenging due to the broad optimization space
and quickly evolving workloads~\cite{DBLP:conf/sosp/WangJZZFNW23,
DBLP:conf/fast/MohanPC21,DBLP:journals/corr/abs-2407-20143}.
For instance, the leading LLM training framework Megatron~\cite{megatron-checkpoint}
has recently integrated concurrent checkpoint optimizations
proposed three years ago~\cite{DBLP:conf/fast/MohanPC21},
with the checkpoint code now comprising a quarter of its codebase.
Meanwhile, many emerging training frameworks, such as those for reinforcement learning~\cite{gym},
still lack these features.
OS-level C/R does not impose any burden on developers for 
implementing and optimizing their own fault tolerance mechanisms.

Concurrent OS-level C/R---performing checkpoint and restore while allowing processes
to run concurrently---is becoming increasingly important,
as stopping GPU processes severely impacts application performance
due to the lengthy data copy time during the C/R process (\textsection{\ref{sec:bg:motiv}}).
For example, Microsoft reports that over 3.9\% of GPU users experience quality issues
from migration stalls caused by C/R~\cite{DBLP:conf/icse/GaoSLZWLY23}.
Moreover,
the 6.2-second stall caused by restoring a Llama2-13B inference using
state-of-the-art OS-level C/R~\cite{DBLP:journals/corr/abs-2202-07848}
is 31$\times$ longer than the Time-To-First-Token (TTFT) for inference,
which significantly hinders deploying GPU processes in emerging serverless computing~\cite{DBLP:journals/corr/abs-1902-03383,ali-serverless-gpu}.
Finally, in model training,
the checkpoint time can be comparable to the iteration time (46--87\%, see \textsection{\ref{sec:eval}}),
and concurrent checkpointing
can save precious GPU hours~\cite{DBLP:conf/sosp/WangJZZFNW23,DBLP:journals/corr/abs-2407-20143,DBLP:conf/fast/MohanPC21,295549}.

\stitle{Current systems and the key challenge. \,}
Existing systems like NVIDIA cuda-checkpoint~\cite{DBLP:journals/corr/abs-2202-07848,nvidia-new-ckpt-driver-repo}
cannot perform concurrent checkpoint and restore of GPU-related states.
While concurrent execution of C/R is standard practice for CPU processes,
it is ineffective when the GPU is stopped because 
the CPU must wait for GPU execution results (\textsection{\ref{sec:bg:motiv}}). 
This raises a key question: 
\emph{Can we achieve efficient and effective OS-level checkpoint and restore during concurrent GPU execution?}

The key challenge
is ensuring correctness---a checkpoint must reflect a valid process state as if no checkpoint had occurred (\textsection{\ref{sec:checkpoint-correct}}).
For example, if the GPU executes two kernels before the checkpoint,
the checkpoint must capture the exact state of GPU memory after these two kernels write.
However, the process running concurrently can corrupt the checkpoint by
overwriting data that has not been checkpointed.
Similarly, during the restore,
if a GPU kernel reads data while the OS is concurrently restoring it,
the kernel may read incorrect data, corrupting the process state.

The key for correctness is to trace the read and write sets of the concurrent execution,
i.e., which bytes of CPU/GPU states are read and written by the process.
Take checkpointing as an example: the OS can use this information
to isolate the writes that cause an incorrect checkpoint
with copy-on-write~\cite{DBLP:conf/sosp/ShapiroSF99,DBLP:conf/sosp/Wu0M023}, or recopy the
writes to the checkpoint~\cite{DBLP:conf/nsdi/ClarkFHHJLPW05} for correctness.
For CPU states,
the OS can leverage OS-mediated data paths with hardware paging to
gather this information,
as summarized in Table~\ref{tab:info}.
However, GPUs lack the necessary hardware support~\cite{DBLP:conf/uss/0002Z024} and bypass the OS during their execution
to maximize performance.

\stitle{Key insights. \,}
Unlike CPU, whose execution is an entire black box to the OS,
the execution of GPU is composed of fine-grained units (e.g., kernels),
whose control flow is mediated by the OS.
Specifically, applications trigger GPU execution through fine-grained GPU API calls (e.g., CUDA~\cite{cuda-driver}).
Moreover, 
each API will trigger fine-grained GPU state modifications (\textsection{\ref{sec:trace-granularity}}),
so the OS can intercept these APIs to trace their read and write sets for concurrent checkpoint and restore. 

However, tracing read and write sets at the API level is non-trivial, 
because although some APIs have well-defined read and write semantics,
e.g., those launch kernels provided by the vendors like \texttt{cuBLAS}~\cite{cuBLAS}, 
the process can trigger user-developed kernels with arbitrary code.
Fortunately, 
we find that although such kernels may include complex code,
each typically serves a clear computational purpose,
such as doing computations on a set of GPU buffers. 
This makes it possible to accurately speculate which data will be accessed 
by simply analyzing the kernel's launch arguments (\textsection{\ref{sec:design-dataflow-track}}).

\newcolumntype{Y}[1]{>{\raggedright\arraybackslash}m{#1}} 

\begin{table}[!t]
    \begin{minipage}{1\linewidth}
        \caption{\small{%
            An overview of information required for correct concurrent checkpoint and restore,
            and how {\sys} traces them.%
        }}
    \label{tab:info}
    \end{minipage} 
    \hspace*{-2mm}
    \begin{minipage}{1.02\linewidth}
    \centering
    \small{
    \resizebox{\columnwidth}{!}{
    \ra{1.1}
    \begin{tabular}{@{~}l l m{3.1cm} Y{1.5cm}@{~}}
        \toprule
        & \textbf{Info.} 
        & \textbf{CPU} 
        & \textbf{GPU} \\
 \hline
 \textbf{Checkpoint} 
        & Writeset 
        & Permission~\cite{DBLP:conf/sosp/Wu0M023,Aurora} and dirty bits~\cite{DBLP:conf/nsdi/ClarkFHHJLPW05} 
		& {Speculation + Validation} \\
 \textbf{Restore}    
        & Readset 
        & Present bits~\cite{DBLP:conf/osdi/WeiLW0Y0023,DBLP:conf/asplos/DuYXZYQWC20,DBLP:conf/eurosys/WangHW19} 
        & \multicolumn{1}{c}{(\textsection{\ref{sec:design-dataflow-track}})} 
        \\ \bottomrule
 \end{tabular}
    }
    } 
\end{minipage} \\[-20pt]
\end{table}  

\stitle{The {\syslong} (\sysshort). \,}
Based on our insights,
we built {\sys}, the first OS service capable of concurrently checkpointing and restoring 
GPU processes~\cite{DBLP:journals/corr/abs-2405-12079}.
{\sys} speculates the data accessed by each GPU kernel using its launch arguments,
and uses this speculated information to retrofit CPU-based concurrent C/R protocols for GPUs, 
including copy-on-write (\textsection{\ref{sec:cow}}), 
recopy (\textsection{\ref{sec:recopy}}), and on-demand restore (\textsection{\ref{sec:design-restore}}).
To ensure correctness, we further instrument a lightweight validator 
to correctly handle speculation failures.

Retrofitting CPU-based concurrent C/R techniques on GPUs with our validated speculation---although ensuring correctness---still
suffers from performance issues 
if GPU features are not considered.
First, during concurrent checkpointing, the interference between
CPU and GPU checkpointing, as well as between application data transfers and checkpoint transfers,
can significantly impact the efficiency of the concurrent checkpointing.
To mitigate this interference,
we design a coordinated and prioritized checkpoint mechanism (\textsection{\ref{sec:checkpoint-fast}}).
Second, before allowing a concurrent restore, we must create the proper GPU execution environment,
which incurs overhead comparable to that of data restoration itself.
We propose using a GPU context pool to
bypass the context creation to fully unleash the power of concurrent restore (\textsection{\ref{sec:design-restore}}).

\begin{figure*}[!t]
            \vspace{1mm}
            \begin{minipage}{1\linewidth}
            \hspace{-1mm}
            \centering    
            \includegraphics[width=1\textwidth, trim=0.25cm 12.3cm 3.2cm 0.25cm, clip]{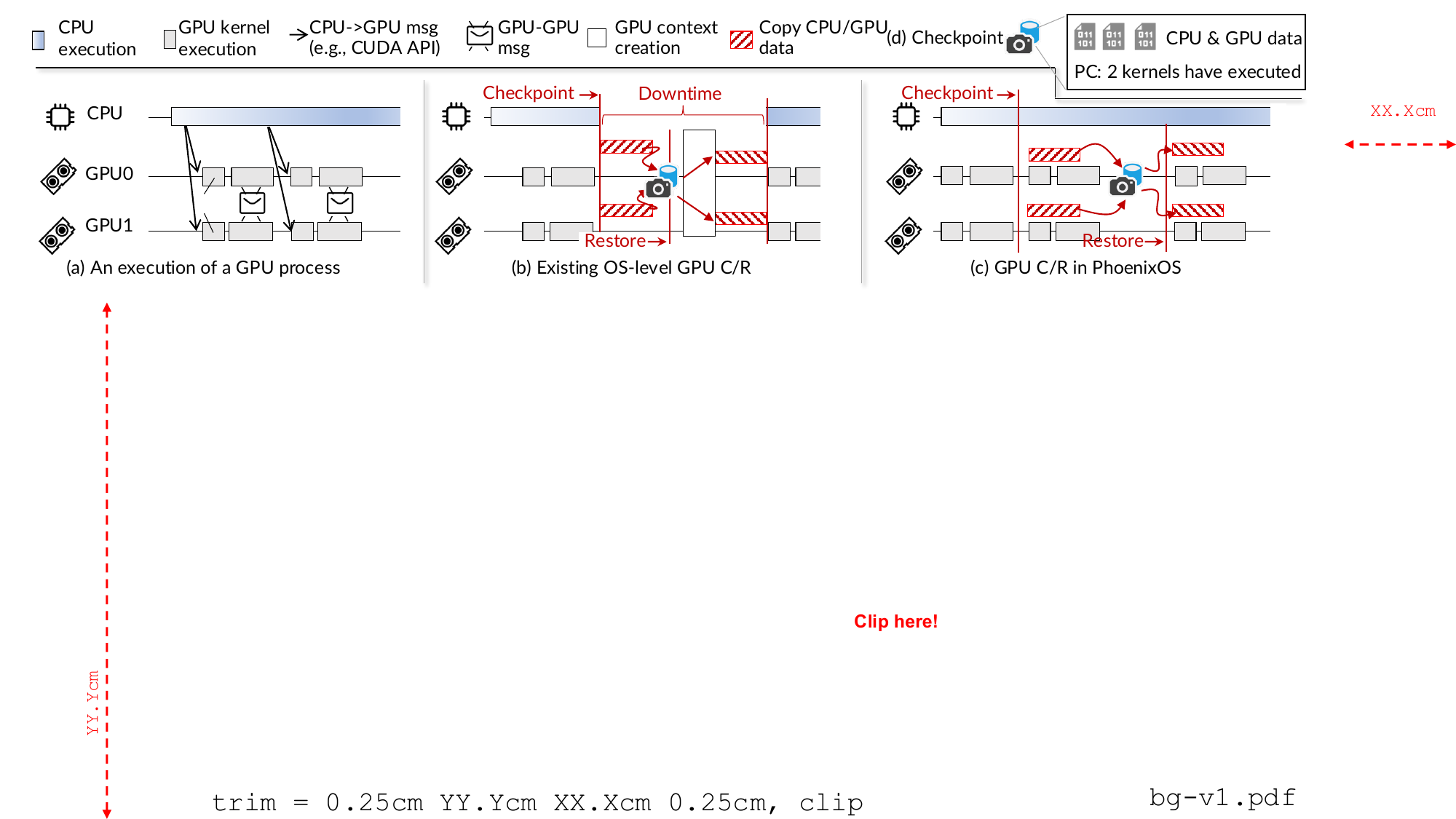}
            \end{minipage} \\[5pt]
            \begin{minipage}{1\linewidth}
            \caption{\small{%
            Illustrations of
            (a) how a GPU process executes, 
            (b) how a stop-the-world OS-level checkpoint and restore works,
            (c) how {\sys} does concurrent checkpoint and restore,
            and (d) the main content of checkpointed data.
            }}
            \label{fig:bg}
            \end{minipage} \\[-5pt]
            \end{figure*}

\stitle{Demonstrations. \,}
{\sys} can checkpoint and restore unmodified GPU applications (including multi-GPU processes) on NVIDIA GPUs.
We selected NVIDIA GPUs for their widespread adoption, 
despite the greater implementation challenges they present.
Our design also generalizes to other GPUs and accelerators that
follow the same execution model of NVIDIA GPUs (\textsection{\ref{sec:bg:gpu}}).
When evaluated on A800 GPU servers and  
compared to state-of-the-art systems---Singularity~\cite{DBLP:journals/corr/abs-2202-07848} 
and NVIDIA's official checkpoint and restore utility 
(cuda-checkpoint)~\cite{nvidia-new-ckpt-driver-repo},
{\sys} reduces application stall during checkpoint and restore by up to 160\%. 
More importantly, {\sys} delivers orders-of-magnitude improvements 
in end-to-end application performance. 
Compared to Singularity,
{\sys} reduces wasted GPU time in fault tolerance by 76\% when training Llama2-13B on multiple GPUs.
It also decreases migration downtime for a Llama2-13B inference job from 9.8 to 2.3 seconds,
Additionally, {\sys} can launch a new Llama2-13B inference job in just 622 milliseconds 
by avoiding cold start costs,
which is 114--342\% and 124--450\% faster than Singularity and cuda-checkpoint, respectively.

\stitle{Contributions. \,} We highlight our contributions as: \\[-12pt]
\begin{itemize}[leftmargin=*,leftmargin=10pt,itemindent=0pt]
    \item The first efficient GPU execution read and write set tracing method
    via validated speculation (\textsection{\ref{sec:design-dataflow-track}}).  \\[-5pt]
    \item The first set of OS-level concurrent checkpoint and restore protocols for GPUs
    (\textsection{\ref{sec:cow}, \textsection{\ref{sec:recopy}}, and \textsection{\ref{sec:design-restore}}}). \\[-5pt]
    \item {\sys}, 
	the first OS-level concurrent checkpoint and restore system for GPUs
    that addresses critical technical challenges to realize these concurrent protocols 
	(\textsection{\ref{sec:checkpoint-fast}} and \textsection{\ref{sec:design-restore}}), 
    significantly improving end-to-end application performance (\textsection{\ref{sec:eval}}).\\[-5pt]
\end{itemize}

\noindent {\sys} is open-source and publicly available at {{\burl{https://github.com/SJTU-IPADS/PhoenixOS}}}.

\section{Background}
\label{sec:bg}

\subsection{How a process uses GPUs}
\label{sec:bg:gpu}

\noindent
GPUs are accelerators with massive multithreading capability
typically attached to the CPU via PCIe.
Programs executed on GPUs are termed \emph{kernels},
which contain machine code (e.g., SASS~\cite{sass})
that is either pre-compiled or just-in-time compiled.
Kernels process data residing in GPU \emph{buffers},
where each buffer constitutes a contiguous GPU virtual memory region
with application-controlled granularity.

When a process starts, 
the GPU driver creates an execution context,
e.g., CUcontext~\cite{cuda-context},
involving compiling and loading the kernel binaries,
configuring the GPU virtual memory and others~\cite{Mai2023HONEYCOMB}.
Afterward,
processes use a GPU driver or toolkit APIs (we term \emph{GPU API} in this paper)
like CUDA~\cite{cuda-driver},
illustrated in {\fig{fig:bg}}(a), to trigger computations on GPUs.
For example, the process can launch kernels via \texttt{{cudaLaunchKernel}}
and copy CPU data to GPU (and vice versa) using \texttt{cudaMemcpy}.
{\sys} intercepts GPU APIs for realizing checkpointing and restoration,
thus supporting all GPU applications at the OS-level.

\subsection{OS-level GPU checkpoint and restore (C/R)}
\label{sec:bg:checkpoint}

\nospacestitle{Basic checkpoint and restore. \,}
A checkpoint captures a process's execution state at a specific time
in an image (termed as \emph{checkpoint}).
The OS can then use this checkpoint to recreate the process (termed as \emph{restore}).
A checkpoint contains two types of data:
data state (data in the CPU memory and GPU buffers) and control state
(e.g., CPU registers that store the program counter, PC).
Checkpointed CPU states also include kernel objects like
network connections~\cite{criu-tcp,Aurora}.

A common approach for implementing GPU checkpointing 
first quiesces (both CPU and GPU) process execution---stopping CPU execution 
and waiting for all in-flight GPU kernels and communications to complete---and then 
copies all execution state to 
the checkpoint~\cite{nvidia-new-ckpt-driver-repo,DBLP:journals/corr/abs-2202-07848},
as illustrated in {\fig{fig:bg}}(b).
For restoration, the OS first loads data state into CPU memory and GPU buffers,
then restores control state to resume execution.

\stitle{Concurrent CPU checkpoint and restore. \,}
To prevent application stall during checkpointing and restoration,
which has non-trivial overhead,
OS researchers have investigated concurrent CPU checkpointing and restoration for decades~\cite{DBLP:conf/sosp/Wu0M023,Aurora,DBLP:conf/nsdi/ClarkFHHJLPW05,DBLP:conf/osdi/WeiLW0Y0023,DBLP:conf/asplos/DuYXZYQWC20,DBLP:conf/eurosys/WangHW19}.
Concurrent checkpointing allows CPU execution during data copying,
thus hiding the overhead of checkpointing.
For correctness,
the OS either isolates concurrent CPU writes via copy-on-write~\cite{DBLP:conf/sosp/Wu0M023,Aurora},
or recopies concurrently written data to the checkpoint~\cite{DBLP:conf/nsdi/ClarkFHHJLPW05}.
Concurrent restoration enables immediate process resumption,
i.e., no need to wait for the data to be fully restored.
During the process execution,
the data is being concurrently copied from the checkpoint to the CPU memory.
For correctness,
if the process touches non-restored data,
the CPU will trigger a page fault,
so the OS will copy the missing data from the checkpoint on demand~\cite{DBLP:conf/osdi/WeiLW0Y0023,DBLP:conf/asplos/DuYXZYQWC20,DBLP:conf/eurosys/WangHW19}.

\subsection{Key factors stalling applications during C/R}
\label{sec:bg:motiv}

\begin{figure}[!t]
        \centering
        \includegraphics[scale=1.1]{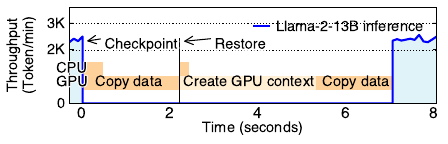} 
        \\[-2pt]
        \begin{minipage}{1\linewidth}
                \caption{\small{                
                A breakdown of checkpoint and restore overhead. 
                }}
        \label{fig:motiv-overall}
    \end{minipage} \\[-15pt]
    \end{figure}  

\noindent
{\fig{fig:motiv-overall}} analyzes the stalls caused by checkpoint and restore
on a Llama2-13B inference process. 
We evaluate Singularity~\cite{DBLP:journals/corr/abs-2202-07848}---the state-of-the-art
GPU checkpoint and restore system,
other baselines like cuda-checkpoint~\cite{nvidia-new-ckpt-driver-repo} are much slower.
We have enabled concurrent CPU checkpointing and restoration 
using CRIU's incremental dump and restore~\cite{criu-memory-change-tracking}.
The detailed setup and performance of cuda-checkpoint can be found in \textsection{\ref{sec:eval}}.

\stitle{Copying GPU data. \,}
Because the GPU is stopped when copying data to and from checkpoint,
the applications are stalled despite the CPU-side of the process being concurrently executed.
The stall time is non-trivial:
checkpointing and restoring data each require more than 2.1 seconds respectively when transferred with PCIe,
causing thousands of tokens to be disrupted.
The stall further scales with the GPU memory used, which is substantial:
For instance, Llama2-13B inference occupies 55\,GB of active memory.
Given a typical 32\,GBps PCIe 4.0 CPU--GPU link,
OS-level checkpoint and restore takes at least 1.7 seconds.

\stitle{Creating GPU contexts. \,}
Before restore can copy data to the GPU,
the OS must create proper GPU contexts.
This establishes a restoration barrier
as context creation incurs comparable overheads to data copying,
i.e., 3.1 vs. 1.7 seconds in our motivation experiment.
GPU context initialization exceeds CPU context creation time
due to complex hardware configuration and driver state loading~\cite{Mai2023HONEYCOMB}.
In comparison, creating CPU context, i.e., a Linux process,
merely initializes a small number of kernel data structures.

\section{{\syslong} ({\sys})} 
\label{sec:overview} 

\nospacestitle{Design goals. \,}
Our goals are \emph{efficient} and \emph{correct} OS-level GPU checkpoint and restore.
For efficiency, the goal is to minimize process stalls during checkpoint or restore.
{\sys} achieves this through:
concurrent process execution during data copying for both checkpoint and restore, and
bypassing GPU context creation during restore,
as illustrated in {\fig{fig:bg}}(c).
For correctness,
the goal is to make the checkpoint reflect an application state
indistinguishable from a non-checkpointed execution~\cite{DBLP:journals/corr/abs-2202-07848,DBLP:conf/ics/LeeSHTKE19,Aurora}.

\begin{figure}[!t]
        \begin{minipage}{1\linewidth}
        \centering    
        \includegraphics[width=1.\linewidth, trim=0.25cm 11.4cm 18.3cm 0.25cm, clip]{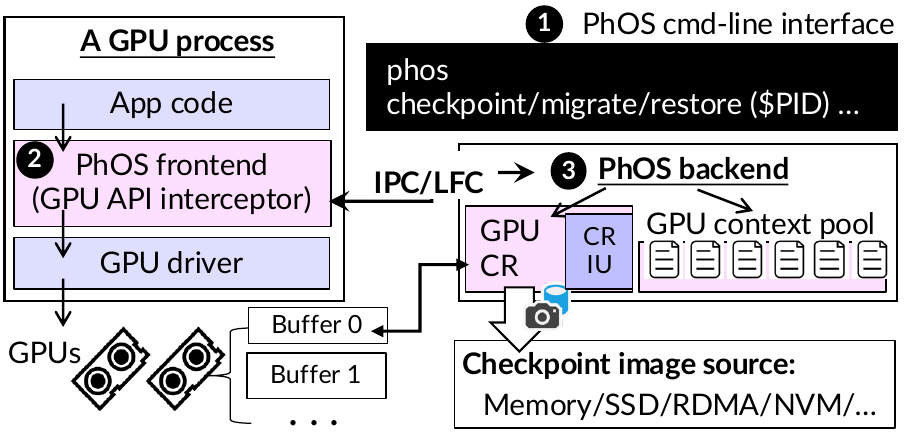}
        \end{minipage} \\[5pt]
        \begin{minipage}{1\linewidth}
        \caption{\small{%
        {\sys} system architecture.
        IPC and LFC stand for inter-process call and local function call, respectively.
        }}
        \label{fig:pos-arch}
        \end{minipage} \\[-15pt]
        \end{figure}

\stitle{System components and execution flow. \,}
{\fig{fig:pos-arch}} shows {\sys}'s architecture
that has three core system components:
a command-line tool (\ding{182}),
a per-process frontend library to facilitate C/R (\ding{183}),
and a backend module to do the C/R (\ding{184}).

Our command-line tool (\ding{182}) enables users to checkpoint running GPU processes,
migrate them across machines, or restore processes from checkpoints
with the process ID (\$PID).
Although {\sys} does not require application cooperation,
we found that it can benefit from choosing a proper checkpoint
time (see \textsection{\ref{sec:eval-timing}}).
We therefore provide an SDK for applications to control checkpoint timing
with just five lines of additional code
(more details in \textsection{\ref{sec:appendix:sdk}}). 

Once a checkpoint or restore command is received by the tool,
it will communicate with the frontend library (\ding{183}) embedded in the GPU driver and its toolkits.
The frontend will further call our backend to do the checkpoint and restore,
as well as trace the read and write sets (\textsection{\ref{sec:design-dataflow-track}})
when the process calls a GPU API.
This information is provided
to our C/R backend to ensure correctness (\textsection{\ref{sec:checkpoint-correct}} and \textsection{\ref{sec:design-restore}}).
Our current implementation leverages \texttt{LD\_PRELOAD}
to extend the CUDA GPU API as CUDA is closed-source,
but \texttt{LD\_PRELOAD} is not a requirement once the code is available.

Our backend (\ding{184})
uses CRIU~\cite{criu} to checkpoint and restore CPU process states,
and our efficient GPU engine (see \textsection{\ref{sec:checkpoint-fast}})
for GPU states.
Choosing CRIU allows {\sys} to correctly checkpoint and restore 
complex states like network connections.
However, CRIU may not be the most efficient solution for CPU checkpointing.
Our current implementation overlaps CRIU checkpointing with our GPU checkpointing
but a faster CPU checkpointing mechanism would make {\sys} easier to implement.

{\sys} backend supports a wide range of checkpoint media:
it can read and write checkpoints to local SSD, CPU DRAM
and even the DRAM of another machine via RDMA (\textsection{\ref{sec:cases}}),
which is critical for efficient process migration.
The backend also includes a GPU context pool---containing pre-allocated contexts
that avoid creating GPU contexts during restore (\textsection{\ref{sec:design-restore}}).

{\sys} command-line tool uses inter-process calls (IPC) to communicate with the frontend.
On the other hand, the frontend and backend can communicate either via IPC or local function calls (LFC).
We prefer LFC whenever feasible by linking the frontend and backend in the same library.
However, IPC is required for cases when the applications
require accessing the context pool,
e.g., for fast restore (see \textsection{\ref{sec:cases}}).
This is because the context must be pre-created in {\sys} daemon to bypass the costly
context creation during restore (see {\fig{fig:motiv-overall}}).
This comes at the cost of extra GPU API calling overhead when the application executes (after the restore),
e.g., we observed a maximum of 9\% IPC overhead for typical AI applications
after applying recent works in accelerating calling GPU APIs
with IPC~\cite{DBLP:journals/corr/abs-2401-13354,DBLP:journals/corr/abs-2306-03622}.
We believe this is a reasonable trade-off as the restore time dominates
the overall application execution time in the required scenarios (see \textsection{\ref{sec:cases}}).

\section{Making concurrent GPU checkpoint correct}
\label{sec:checkpoint-correct}

\nospacestitle{Correctness guarantee of {\sys}. \,}
Intuitively, a checkpoint is correct
if it could occur in a checkpoint-free execution~\cite{DBLP:journals/csur/ElnozahyAWJ02}.
A stop-the-world checkpoint described in \textsection{\ref{sec:bg:checkpoint}}
naturally guarantees correctness.
\emph{{\sys} ensures correctness by ensuring that our checkpoint
 matches the one that may come from the existing stop-the-world checkpoint.}

 \begin{figure}[!t]
        \begin{minipage}{1\linewidth}
        \centering    
        \includegraphics[width=1\linewidth, trim=0.25cm 12.2cm 15.2cm 0.25cm, clip]{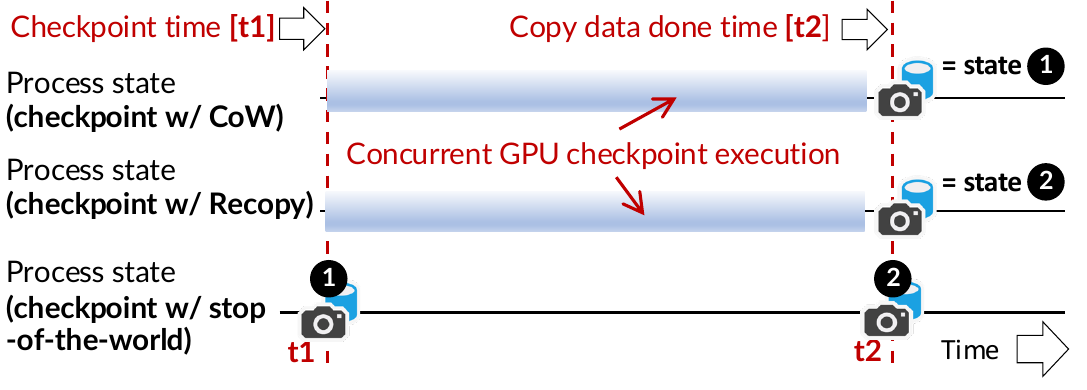}
        \end{minipage} 
        \begin{minipage}{1\linewidth}
        \caption{\small{%
        An illustration of the correctness guarantee of concurrent checkpointing
        enabled by different protocols of {\sys}. 
        }}
        \label{fig:checkpoint-correctness}
        \end{minipage} \\[-15pt]
        \end{figure} 

\stitle{Overview of the checkpoint protocols. \,}
We retrofit
two CPU-inspired protocols~\cite{DBLP:conf/sosp/Wu0M023,Aurora,DBLP:conf/nsdi/ClarkFHHJLPW05}
for a correct concurrent GPU checkpoint.
As {\fig{fig:checkpoint-correctness}} shows,
consider our checkpoint starting at \texttt{t1} with concurrent data copy done at \texttt{t2}.
Compared with a stop checkpoint at \texttt{t1},
inconsistencies are caused by new writes issued between \texttt{[t1,t2]}.
Our soft copy-on-write (\emph{CoW}) protocol isolates these writes (\textsection{\ref{sec:cow}}),
making the checkpoint only copy a frozen GPU state at \texttt{t1}.
Alternatively, compared to a stop checkpoint at \texttt{t2},
our approach may miss new writes.
Thus, at \texttt{t2},
our soft recopy protocol
quiesces the process and recopies the missed writes to ensure correctness (\textsection{\ref{sec:recopy}}).
CoW is commonly used in fault tolerance cases,
where the application can tolerate restoring from a stale state (\texttt{t1}).
On the other hand,
recopy is used in cases where the restored process
must be resumed from the latest execution (\texttt{t2}),
e.g., live migration (\textsection{\ref{sec:cases}}).

Both protocols require tracing the process's \emph{write set}, 
i.e., which bytes are written during the concurrent checkpoint.
For CPU writes, we follow the traditional approach to use permission bits in the page table,
i.e., write protection for copy-on-write
and dirty bits to trace at the page granularity. 
We omit the detailed CPU-side description since it is well studied.
Unfortunately, modern GPUs lack these supports, 
e.g., no hardware dirty bits~\cite{DBLP:conf/uss/0002Z024}.
Therefore, before delving into our protocols,
we first propose a software-based approach to trace GPU writes.

\subsection{Tracing GPU write set with validated speculation}  
\label{sec:design-dataflow-track}

\nospacestitle{Tracing with speculation on GPU API arguments. \,} 
All operations that can perform GPU writes are initiated through
GPU API calls, so we intercept these calls to trace the write set.
These APIs fall into four categories:  \\[-12pt]
\begin{enumerate}[leftmargin=*,leftmargin=15pt,itemindent=0pt]
    \item Issue memory move operations, e.g., 
    \texttt{cudaMemcpy(dst,\\src,count,..)}. 
    They write the CPU data (\texttt{src}) to the GPU memory (\texttt{dst}).
    \\[-8pt]

    \item Issue communication kernels, e.g., 
    \texttt{ncclBroadcast\\(sendbuffer,recvbuffer,count,..)}. 
    They write the GPU buffers with the content sent through the network. 
    \\[-8pt]        

    \item Issue computation kernels with well-defined semantics, 
    e.g., \texttt{cublasSgemm(..,A,B,..,C,..)}. 
    They perform a computation and update the result buffers (\texttt{C}), 
    whose read and write semantics are known in the specifications~\cite{cuBLAS}. 
    \\[-8pt]
    
    \item Issue opaque computation kernels, e.g.,
    \texttt{cudaLaunch-\\Kernel(func,..,args,..)}.
    They launch kernels written by the developers
    or just-in-time compiled during runtime,
    so the reads and writes are unknown to the OS.
    \\[-10pt]
\end{enumerate}

\noindent For types 1--3 APIs,
{\sys} directly uses their specifications to trace writes.
For instance, the specification of \texttt{cudaMemcpy} indicates that executing this API
will modify GPU memory from \texttt{dst} to \texttt{dst+count}.
Our empirical analysis shows that 
over 50\% of invocations are these types of APIs (\textsection{\ref{sec:eval-factor}}).

\begin{figure}[!t]
        \begin{minipage}{1\linewidth}
        \centering    
        \includegraphics[width=1\linewidth, trim=0.25cm 13.9cm 19.0cm 0.25cm, clip]{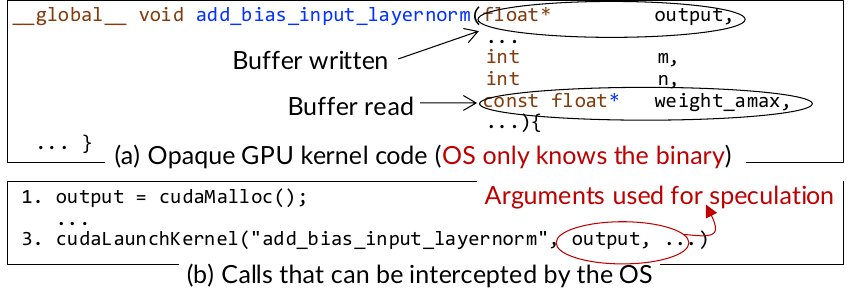} 
        \end{minipage} \\[3pt]
        \begin{minipage}{1\linewidth}
        \caption{\small{%
        An illustration of the GPU kernel~\cite{fasttransformercode} written by the user (a)
        and how {\sys} speculates the accessed buffers (b). 
        }}
        \label{fig:spec-example}
        \end{minipage} \\[-15pt]
        \end{figure} 

Opaque kernels are trickier because
the OS only knows the kernel binary and the arguments invoking it.
Our observation is that each GPU kernel represents 
a clear and straightforward computational purpose
whose accessed data addresses are directly encoded in the OS-known
arguments calling the kernel.
As shown in {\fig{fig:spec-example}}, 
the data (\texttt{output}) written by the kernel
is the second argument.
Since the OS knows all the buffers 
allocated by the process
by intercepting the GPU memory allocation calls, line 1 (\texttt{cudaMalloc}),
and it knows each kernel's function signature~\cite{cuda-signature}, 
we can systematically compare the arguments
with the allocated buffers to speculate the buffer written by each kernel.
The interception overhead is negligible:
it only manipulates a few data structures before calling GPU
memory allocation APIs.

Specifically, our speculation traces the kernel's writes at \emph{buffer-level}:
When applications launch an opaque GPU computation kernel,
for each argument, {\sys} treats it as a tentative address pointing to a GPU buffer to write. 
To decide which buffer is written, 
we compare it with all the process's allocated buffers. 
If the integer value of the argument falls within the range of a buffer,
we mark the entire buffer as written.
To reduce false positives,
we filter out irrelevant arguments with the function signature.
Specifically, we use clang~\cite{clang} to extract the kernel's argument types,
focusing solely on mutable pointer arguments.
One type that cannot be precisely filtered out is
the C struct type, which is opaque to {\sys} as we don't know the struct definition.
For such a type,
we conservatively treat
all 8-byte chunks in the struct as potential written GPU buffers.

\stitle{Runtime validation. \,} 
To handle speculation failures, e.g., extremely rare GPU indirect access (see \textsection{\ref{sec:eval-speculation}}),
we validate the speculation by instrumenting a runtime
validator to the kernel.
The validator checks whether each kernel's GPU memory write instruction
falls within the speculated buffers. 

\begin{figure}[!t]
        \vspace{-2mm}
        \begin{minipage}{1\linewidth}
        \hspace{-2mm}
        \includegraphics[width=1.\linewidth, trim=0.25cm 16.15cm 18.2cm 0.25cm, clip]{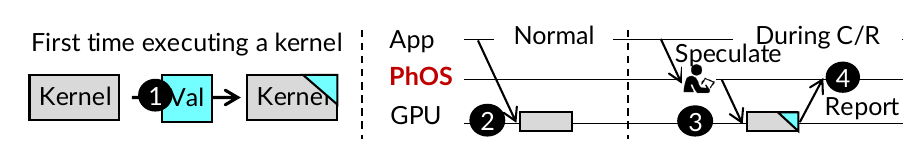} 
        \end{minipage} \\[0pt]
        \begin{minipage}{1\linewidth}
        \caption{\small{%
            {\sys} validation workflow.
        }}
        \label{fig:validation}
        \end{minipage} \\[-20pt]
        \end{figure}

{\fig{fig:validation}} presents the validation workflow.
When {\sys} encounters an opaque kernel that has not been instrumented
(including JIT-compiled ones~\cite{torch-jit}),
it generates a new twin kernel whose kernel binary is the instrumented version
of the original kernel binary (\ding{182}).
The instrumented validator performs the following:
For each write instruction,
it inserts address range checks before it
to verify target address belongs to speculated buffers.
If the validation fails,
the validator reports the incident to {\sys}
by writing the address to a pre-allocated {\sys} managed CPU buffer.
The instrumentation is at the PTX ISA~\cite{ptx}-level for portability.
The instrumentation overhead is negligible because for each kernel binary,
it will be instrumented only once. 

When {\sys} intercepts a kernel invocation during checkpoint (\ding{184}),
it invokes its corresponding instrumented twin kernel. 
If the kernel reports validation failures,
we will execute fallback protocols (described in \textsection{\ref{sec:cow}} and \textsection{\ref{sec:recopy}}) 
to ensure correctness.
The overhead of running the instrumented kernels is also manageable (\textsection{\ref{sec:eval-factor}})
and they are not invoked without a checkpoint (\ding{183}).

\stitle{Discussion: the granularity of tracing. \,}
Ideally, the trace should be as fine-grained as possible
to avoid protocol overhead for ensuring correctness,
e.g., avoid excessive copy-on-write.
For non-opaque kernels,
we can precisely track the bytes written.
However, our speculation-based tracing can only detect writes at the buffer-level,
which may result in over-tracing.
For example,
when a kernel only writes a small part of the buffer,
we may treat the entire buffer as written.
Luckily, such over-tracing is rare, especially for recent GPU applications like AI jobs,
because (1) AI frameworks (e.g., PyTorch)
use a fine-grained buffer allocation,
e.g., one for each tensor~\cite{pytorch-mem},
and (2) the opaque kernels typically write the entire buffer (tensor).
See \textsection{\ref{sec:trace-granularity}} for the analysis. 

\subsection{The GPU soft copy-on-write (CoW) checkpoint}
\label{sec:cow}

\noindent
Suppose a checkpoint starts at time \texttt{t1}.
Our CoW protocol guarantees the final checkpoint matches
a stop-the-world checkpoint at \texttt{t1},
while allowing concurrent application execution.
The inconsistency can only happen when 
the GPU writes to an uncheckpointed buffer during execution.
To prevent this, 
before allowing such writes to execute,
we first copy the targeted data to another place (e.g., a free on-device buffer) and
then redirect application writes to a new buffer.
Thus, the concurrent checkpoint will not see the new writes.

\begin{figure}[!t]
        \begin{minipage}{1\linewidth}
        \centering    
        \includegraphics[width=1\linewidth, trim=0.27cm 12.7cm 16.9cm 0.25cm, clip]{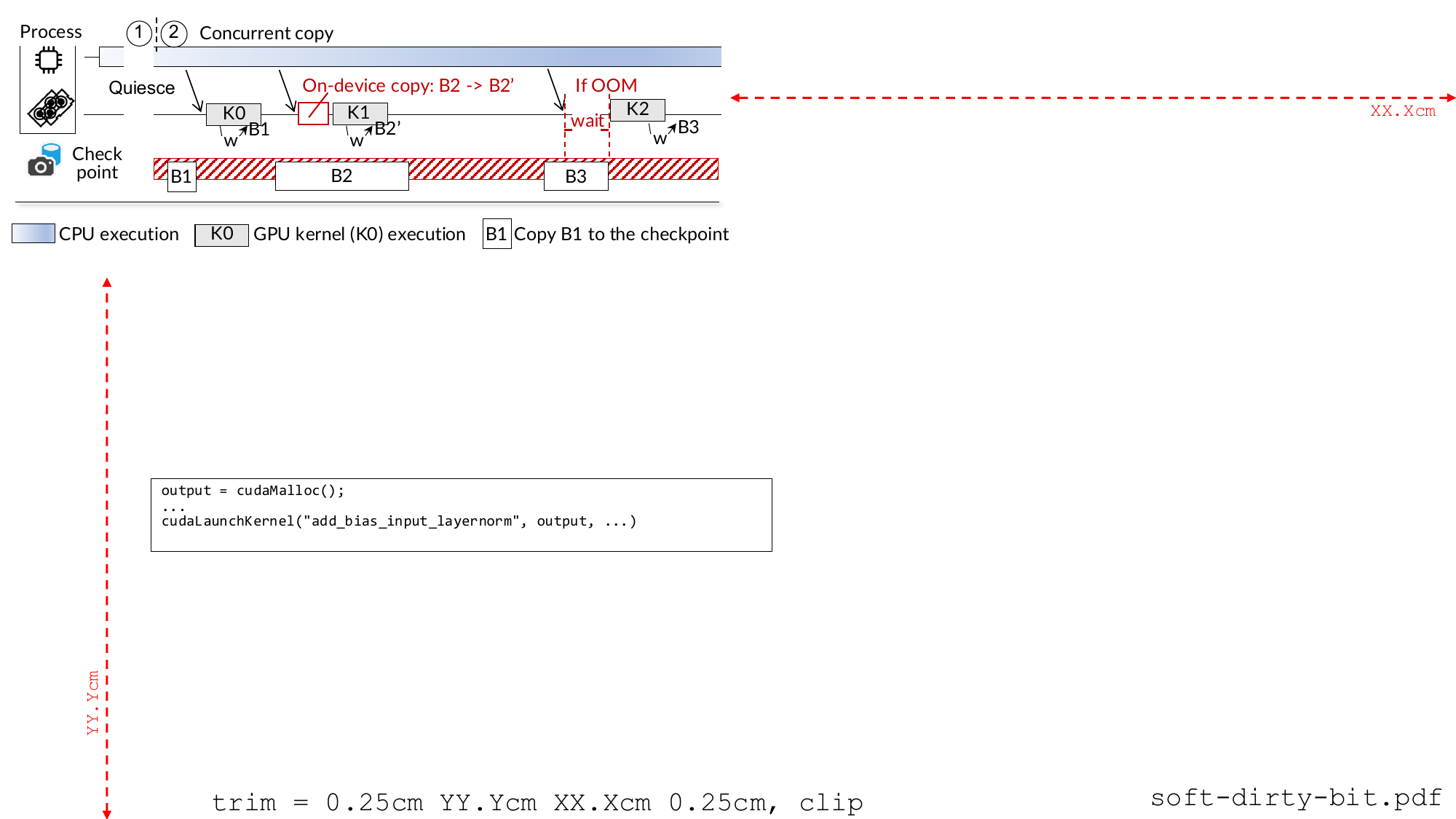} 
        \end{minipage}  \\[-2pt]
        \begin{minipage}{1\linewidth}
        \caption{\small{%
        An illustration of the soft copy-on-write protocol
        for concurrent checkpoint correctness. 
        }}
        \label{fig:soft-cow}
        \end{minipage} \\[-20pt]
        \end{figure}

\stitle{The protocol.\,}
{\fig{fig:soft-cow}} presents the detailed protocol that
consists of two phases: quiesce (\ding{192}) and concurrent copy (\ding{193}).
The quiesce phase stops CPU and GPU execution,
the same as existing stop-the-world checkpointing~\cite{DBLP:journals/corr/abs-2202-07848,nvidia-new-ckpt-driver-repo}.
Note that if we want to checkpoint a multi-process application,
the quiesce phase stops all the processes' CPUs and GPUs.
Quiescing is necessary because it regulates the current process states
to be the same as a possible stop-the-world checkpoint.
To quiesce,
we first stop CPU to prevent sending new GPU APIs.
Then we wait for pending GPU kernels and communications to complete (e.g., via \texttt{cudaDeviceSynchronize}).
Note that the quiescing overhead is negligible since GPU operations occur at microsecond scales.
Thus, despite lacking concurrency, its overhead is significantly lower than data copying
 (recall {\fig{fig:motiv-overall}}).

At the copy phase, we copy all the GPU and CPU data to the checkpoint.
For CPU data, we follow existing works~\cite{DBLP:conf/sosp/Wu0M023,Aurora} 
to isolate concurrent writes with OS's copy-on-write. 
For GPU, 
we isolate writes at the GPU API-level:
Before executing a GPU kernel (or API), 
we check whether it writes to a buffer that has not been checkpointed
(or is being concurrently checkpointed)
using the traced writes through argument-based speculation. 
If so (e.g., B2), we will first copy the victim buffer to a new buffer allocated on the GPU (B2'),
and then redirect the kernel to write to the new buffer (B2').
If multiple concurrent kernels try to write B2,
the first kernel will copy B2 to B2',
and all the others will wait for the copy to finish before executing on B2'.
If the GPU has no free memory to allocate the new buffer,
we will directly copy the data to the checkpoint at the host. 

\stitle{Kernel buffer redirection, on-device memory used, and handling mis-speculation. \,}
Three things need to be noted. 
First, redirecting kernel writes to the new buffer is not that trivial:
For types 1--3 APIs, we can directly change the arguments for so;
but for the opaque ones, we cannot because the arguments can be mis-speculated.
For example, if the changed argument is not a buffer,
the execution will corrupt both application states as well as the checkpoint.
Hence, 
unlike traditional CPU copy-on-write,
we leave the kernel argument unchanged, 
and redirect the checkpoint to access the copied buffer instead. 

Second, we only reserve a small amount of GPU memory (up to 2\,GB) for the copy-on-write,
and the concurrent checkpoint can still work even if no extra memory is used.
If we lack sufficient GPU memory for the copy-on-write,
we block the kernel until free memory is available (K2 in {\fig{fig:soft-cow}})
or the written buffer has been checkpointed. 
This causes a short checkpoint stall but we have found it acceptable:
{\sys} does not require a large on-device buffer for high performance because
once a copy-on-write is done, its original buffer can be immediately released.
\textsection{\ref{sec:eval-mem-capacity}} analyzes the impact of GPU memory reserved
for CoW on the checkpointing performance.

Finally, if the validator detects a mis-speculation by reporting a write to a buffer
not traced,
we will check whether it affects correctness and react if necessary.
If {\sys} has checkpointed the buffer before kernel execution,
no further action is required.
Otherwise, we discard the current checkpoint and retry with a stop-the-world approach
for liveness.
Note that because we don't encounter speculation failures for all major GPU applications,
we adopted a simple retry strategy and leave a more advanced one as our future work.

\stitle{Correctness. \,}
Our checkpoint is the same as
the resulting checkpoint from a stop-the-world checkpoint
at the start checkpoint time.
First, our quiesce phase
is the same as the stop checkpoint.
So before the copy phase, our initial process states are the same as a stop checkpoint.
Second, during our concurrent copy, we only copy the states not modified by concurrent execution,
because all writes are isolated by copy-on-write.
As the stop checkpoint also faithfully checkpoints the same state,
our checkpoint is the same as its checkpoint.

\stitle{Discussion: speculation vs. validation. \,}
Although we can, in principle, avoid speculation by
instrumenting the kernels to obtain their accessed buffers
and do the copy-on-write, as we did in validation,
we still choose speculation for two reasons.
First, speculation allows us to acquire buffer information \emph{in advance},
which is necessary for CoW especially in case of insufficient on-device buffer.
Also, our concurrent restore protocol (see \textsection{\ref{sec:design-restore}})
relies on knowing the accessed buffers in advance.
Second, speculation then validation has fewer overheads than
obtaining the buffers without speculation,
because in the common case, the validation will only pass a flag indicating
whether the validation passes.

\subsection{The GPU soft recopy checkpoint} 
\label{sec:recopy}

\begin{figure}[!t]
        \begin{minipage}{1\linewidth}
        \centering    
        \includegraphics[width=1\linewidth, trim=0.27cm 12.7cm 16.9cm 0.25cm, clip]{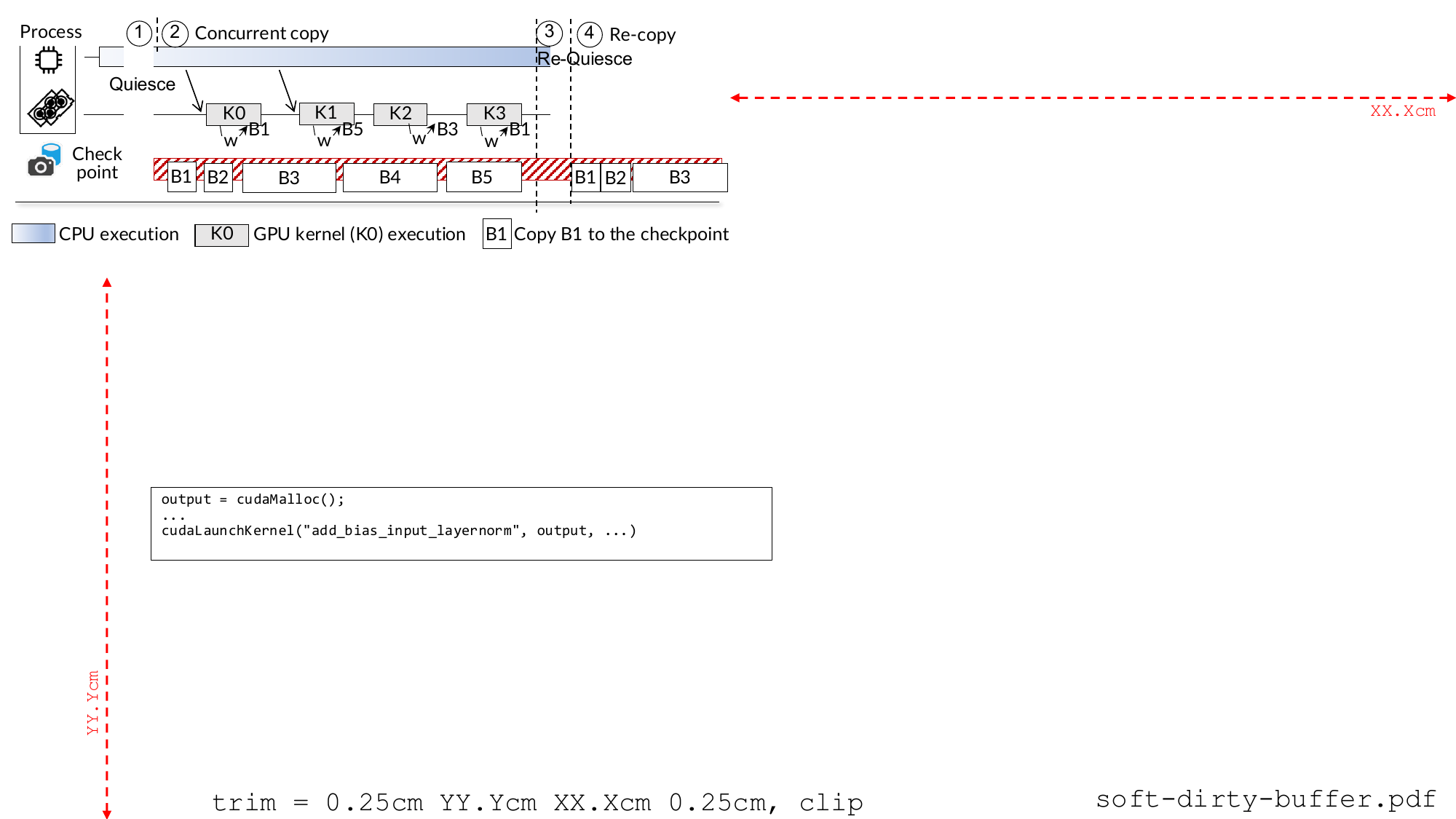} 
        \end{minipage}  \\[-2pt]
        \begin{minipage}{1\linewidth}
        \caption{\small{%
        An illustration of the soft dirty buffer protocol
        for concurrent checkpoint correctness. 
        }}
        \label{fig:soft-dirty-buffer}
        \end{minipage} \\[-20pt]
        \end{figure}

\noindent
Suppose the concurrent checkpoint starts at \texttt{t1} and the data copy ends at \texttt{t2}.
Our recopy protocol recopies the written buffers during concurrent copy 
to ensure the checkpoint is the same 
as the stop-the-world checkpoint at \texttt{t2}.

\stitle{The protocol. \,}
{\fig{fig:soft-dirty-buffer}} presents how the protocol executes,
which contains four phases. 
The first quiesce phase (\ding{192}) stops all the CPU and GPU execution,
the same as the CoW protocol,
ensuring that during concurrent copy, we will not miss tracing any writes.
Afterward, {\sys} resumes application execution, and copies the 
CPU and GPU data to the checkpoint (\ding{193}).
During the concurrent copy,
before launching a kernel (or an API),
we trace the buffers written by the kernel using the speculation and validation method.
For each buffer, we will check whether it is dirty, 
i.e., the buffer has been checkpointed or is being checkpointed.
If dirty, we will add it to a dirty buffer set for recopy.
In the {\fig{fig:soft-dirty-buffer}} example,
B1 and B3 are dirty but B5 is not.
The CPU dirty pages are tracked via page table's dirty bit. 

Once all the data has been copied, we start another quiesce phase
to stop the CPU and GPU execution (\ding{194}).
Note that dirty buffers written during the quiesce phase will also be recorded.
After the quiesce is done,
we recopy all the GPU dirty buffers CPU dirty pages to the checkpoint (\ding{195}).
The application is stopped to ensure correctness,
but we can also iteratively do the concurrent recopy similar to CPU-based protocols~\cite{DBLP:conf/nsdi/ClarkFHHJLPW05}.

\stitle{Handling mis-speculation. \,}
Handling speculation failures is simple for recopy: 
if the speculation is wrong, the validator will
return the victim buffers such that we can add them to the dirty buffer set
if they are dirty. 

\stitle{Correctness. \,}
Assume the concurrent copy completes at \texttt{t2}.
We will show that our checkpoint
 matches a stop checkpoint at \texttt{t2}.
First, it is straightforward to see that the re-quiesce and recopy phases (\ding{194} + \ding{195})
are nearly identical to a stop checkpoint at \texttt{t2}.
The only difference is that we only copy the dirty buffers.
To remedy, we use the buffers copied during \ding{193}
for the remaining buffers.
It is correct to do so because buffers not recopied
 are the same as when doing a stop checkpoint
at \texttt{t2} because they are not dirty.
Thus, combining buffers copied from the two phases results in the same checkpoint
as obtained from a stop checkpoint at \texttt{t2}.

\section{Making concurrent GPU checkpoint fast}
\label{sec:checkpoint-fast}


\begin{figure}[!t]
        \begin{minipage}{1\linewidth}
        \centering    
        \includegraphics[width=1\linewidth, trim=0.27cm 14.5cm 19.6cm 0.25cm, clip]{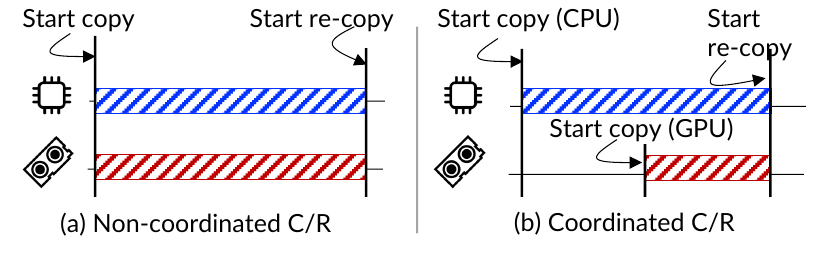} 
        \end{minipage}  \\[3pt]
        \begin{minipage}{1\linewidth}
        \caption{\small{%
        An illustration of the reduced time window size
        that could cause dirty buffers thanks to
        prioritized CPU and GPU concurrent checkpoint.
        The bars show the time window size. 
        }}
        \label{fig:conflict}
        \end{minipage} \\[-10pt]
        \end{figure} 

\nospacestitle{Reducing dirty buffers via coordinated GPU and CPU checkpoint. \,}
In the soft recopy protocol,
the time window of the copy is critical to the number of dirty buffers generated:
the longer the copy, the more likely the buffers will become dirty.
If we copy the CPU and GPU data without coordination, 
the overall time window is the size of checkpoint image divided by the copy bandwidth.

In GPU processes, we found GPU writes are more frequent than CPU writes,
so it is beneficial to first copy the CPU data instead of copying the GPU data,
reducing the GPU time window to the size of the GPU data divided by the copy bandwidth,
as shown in {\fig{fig:conflict}}(b).
Thus, we implement a coordinated checkpoint in our checkpoint engine:
we first copy the CPU data, then the GPU data.

\stitle{Prioritized application PCIe transfer. \,}
When we completed our first concurrent checkpoint implementation (for both protocols),
we found the process still has significant stalls during the checkpoint.
By profiling the GPU execution,
we found the stall is not caused by our protocols,
but by the saturated DMA engine caused by GPU checkpointing.
Specifically,
GPUs have a limited number of DMA engines~\cite{bakitademystifying}
shared between {\sys} and applications.
With concurrent checkpointing,
the bulk copy nature of the checkpointing can easily saturate the DMA engines,
causing application kernels to wait.

To this end, we adopt two techniques to avoid application starvation---the goal is to 
prioritize application DMA transfer
over the checkpoint, observing that applications infrequently use
DMA but DMA operations are on the critical path.
First, we intercept all process's DMA-related API calls (e.g., \texttt{cudaMemcpy})
to detect application PCIe transfer.
The interception overhead is negligible (<1\%),
because the APIs are typically called asynchronously 
so our interception is not on the application's critical path~\cite{DBLP:journals/corr/abs-2401-13354,DBLP:journals/corr/abs-2306-03622}.
When the DMA operation is detected, 
we implemented a preemptible checkpoint copy mechanism: 
for each GPU buffer, we copy it to the checkpoint in small 4\,MB chunks.
After copying one chunk, we check whether there is ongoing or pending application transfer.
If so, we pause the checkpoint copy and let the pending application transfers execute first.

\section{Concurrent GPU restore}
\label{sec:design-restore}

\begin{figure}[!t]
        \begin{minipage}{1\linewidth}
        \centering    
        \includegraphics[width=1\linewidth, trim=0.27cm 14cm 17.7cm 0.25cm, clip]{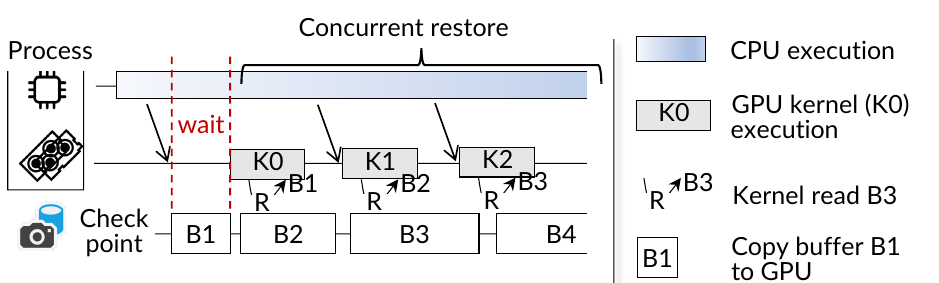} 
    \end{minipage}  \\[5pt]
        \begin{minipage}{1\linewidth}
        \caption{\small{%
        An illustration of the concurrent restore in {\sys}. 
        }}
        \label{fig:concurrent-restore}
        \end{minipage} \\[-20pt]
        \end{figure}

\nospacestitle{Concurrent and correct restore. \,}
When restoring data from the checkpoint to the GPUs,
{\sys} allows processes to execute concurrently
when {\sys} is copying data from the checkpoint to the GPU buffers.
{\fig{fig:concurrent-restore}} illustrates the execution flow.
To ensure correctness,
before executing a kernel (or API),
we check whether the data it requires has been copied.
If not, we pause the execution, copy the buffer containing the required data,
and then resume.
In our example,
K0 will wait for B1 to be copied.
If the data has been copied,
e.g., K1 and K2,
kernels can execute while the restore is concurrently executing (copying B3 and B4).
This effectively overlaps computation with data transfer,
reducing the observable impact of data copy overhead analyzed in \textsection{\ref{sec:bg:motiv}}.

Recall from \textsection{\ref{sec:checkpoint-correct}} that the checkpoint of {\sys} is correct.
Given this guarantee,
the correctness of concurrent restore depends on accurately tracing execution's read set,
similar to prior CPU systems~\cite{DBLP:conf/osdi/WeiLW0Y0023,DBLP:conf/asplos/DuYXZYQWC20,DBLP:conf/eurosys/WangHW19}.

\stitle{Tracing GPU read set with extended speculation with validation. \,}
{\sys} extends the argument-based speculation described in \textsection{\ref{sec:design-dataflow-track}}
to speculate the data read by launching a piece of GPU execution with API calls.
The extension is simple: we treat launch arguments declared as immutable pointers (e.g., \texttt{const void *})
as tentative read buffers required.
Note that we still need to trace writes here,
because GPU kernels may perform partial writes to buffers.

\stitle{Handling mis-speculation. \,}
Like writes, we also instrument a validator to ensure correctness under speculation failures.
If the speculation is wrong, the validator will notify {\sys} to handle.
However, the mis-speculation handling for reads is more complex than for writes,
because the kernel may have accessed inconsistent GPU states caused by a partially restored buffer.
To ensure correctness, we must roll back the GPU states to a correct one.
Our current solution rolls back to the initial state from the checkpoint
and then performs a stop-the-world restore for liveness.
We choose this simple solution because we have not met speculation failures in common GPU applications.
While a more efficient solution could involve isolating partial updates using our soft copy-on-write protocol,
we defer the detailed design to future work.

\stitle{Accelerate restore with context pool. \,}
The final challenge for implementing concurrent restore 
is that GPU context creation must precede kernel execution and buffer copying, 
as the context initialization involves setting up GPU memory subsystems. 
This initialization incurs comparable overhead to data copy, 
as analyzed in \textsection{\ref{sec:bg:motiv}}.

To address the issue, our key observation is that
GPU execution can be decoupled from the context it uses:
a process can use any context capable of executing the APIs.
Therefore, we pre-create a pool of common GPU contexts at {\sys}.
Specifically, we maintain the pool in the {\sys} daemon---a
long-running OS service that pre-creates
CUDA and cuBLAS contexts with \texttt{cuCtxCreate} and \texttt{cublasCreate} at boot time.
For multi-GPU applications, we also pre-create an NCCL group communicator that covers
all GPUs connected via NVLink, which can accelerate establishing the NCCL group communicator
with sub-topology using \texttt{ncclCommSplit}~\cite{ncclgroup} efficiently.
We don't create communicators across machines because the ideal communicator may depend
on the network topology.
The context pool operates in a separate process space,
so applications must use inter-process communication mechanisms to access it.

With the context pool,
when a GPU process requests to create a context,
we intercept the context creation API call and assign a pre-existing context from the pool.
We also track the mapping between the context and the GPU process,
ensuring all subsequent GPU API calls from that process utilize the mapped context.

\section{Empowering applications with {\sys}} 
\label{sec:cases}

\nospacestitle{Fault tolerance for GPU processes (checkpoint-mostly). \,}
Distributed computing applications such as training models are susceptible to GPU failures~\cite{DBLP:conf/icse/GaoSLZWLY23,DBLP:conf/nsdi/EisenmanMIMKNSA22,DBLP:journals/corr/abs-2202-07848}.
{\sys} provides transparent fault tolerance through periodic GPU process checkpointing 
using our soft copy-on-write (CoW) protocol described in \textsection{\ref{sec:cow}}. 
The checkpoints are stored in fast fault-tolerant storage,
e.g., with replication to remote memory~\cite{DBLP:conf/sosp/WangJZZFNW23}.
If a failure occurs, {\sys} first stops all GPU processes,
and then restores them from the most recent checkpoint. 
The checkpoint frequency is determined by holistically considering both
the checkpoint overhead as well as the loss of computation due to restoring from a
stale checkpoint, similar to prior work~\cite{DBLP:conf/eurosys/GuptaKKVGKRS24}.
The detailed method is left in the appendix (\textsection{\ref{sec:appendix:freq}}).
For fault tolerance, using a slightly stale checkpoint (i.e., one that does not capture the absolute latest execution state)
is acceptable, because the latest state can be recovered through recomputation. 

Two things to note.
First, fault tolerance is a checkpoint-mostly case,
as the checkpoint frequency needs to be high to avoid losing computation results~\cite{DBLP:conf/sosp/WangJZZFNW23}.
In comparison, the restore only happens upon failures,
which occur much less frequently, e.g., one per-hour~\cite{DBLP:conf/fast/MohanPC21,DBLP:journals/corr/abs-2205-01068}.
Second, we need to ensure the checkpoint from all the involved processes is consistent~\cite{DBLP:journals/corr/abs-2202-07848}.
Thus, we extended the quiescing phase described in \textsection{\ref{sec:cow}}
across all involved processes.
After the quiescent point, we can checkpoint each process with CoW separately.
We currently follow Singularity~\cite{DBLP:journals/corr/abs-2202-07848} and
use a user-provided hint for a correct global quiesce,
e.g., before the forward pass of each training iteration.

\begin{table*}[!t]
    \vspace{1mm}
    \begin{minipage}{1\linewidth}
    \caption{\small{%
        Applications evaluated.
        ($^{\star}$) indicates a multi-GPU setup.
        PPO is training-only, and our testbed cannot run L$^{\text{70B}}$ 
		training.%
        }}
    \label{tab:app-data}
    \end{minipage} 
    \begin{minipage}{1\linewidth}
    \centering
    \small{
            \ra{1.1}
            \begin{tabular}{@{}l|cc|cc|cc|cc|cc@{}}
                \toprule
                & \multicolumn{2}{c|}{\textbf{ResNET-152M}}
                & \multicolumn{2}{c|}{\textbf{PPO-336M}}
                & \multicolumn{2}{c|}{\textbf{Stable-Diffusion-1B}}
                & \multicolumn{2}{c|}{\textbf{Llama2-13B}} 
                & \multicolumn{2}{c}{\textbf{Llama3.3-70B}} \\ 
                & \multicolumn{2}{c|}{(R$^{\text{152M}}$)}
                & \multicolumn{2}{c|}{(PPO$^{\text{336M}}$)}
                & \multicolumn{2}{c|}{(SD$^{\text{1B}}$)}
                & \multicolumn{2}{c|}{(L$^{\text{13B}}$)} 
                & \multicolumn{2}{c}{(L$^{\text{70B}}$)} \\ 
                \midrule
                \textbf{Application}
                & \textbf{Training} & \textbf{Inference}               
                & \multicolumn{2}{c|}{\textbf{Training}}               
                & \textbf{Training}$^{\star}$ & \textbf{Inference}     
                & \textbf{Training}$^{\star}$ & \textbf{Inference}     
                & \textbf{} & \textbf{Inference}$^{\star}$ \\  
                \textbf{Library}
                & \multicolumn{2}{c|}{Torchvision~\cite{torchvision} } 
                & \multicolumn{2}{c|}{OpenAI Gym~\cite{gym}}            
                & \multicolumn{2}{c|}{HuggingFace~\cite{huggingface}}   
                & \multicolumn{2}{c|}{Meta Llama 2~\cite{llama}}        
                & \multicolumn{2}{c}{Meta Llama 3.3~\cite{llama3}} \\     
                \bottomrule
            \end{tabular}
    }
    \end{minipage} \\[-5pt]
\end{table*}

\begin{figure*}[!t]
    \centering
    \includegraphics[left, scale=1.2]{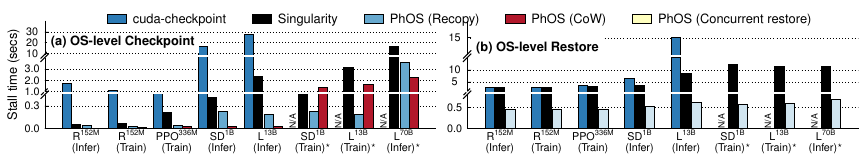} \\[-5pt]
    \begin{minipage}{1\linewidth}
        \caption{\small{
            The application stall time caused by different OS-level checkpoint and restore systems.
        }}
    \label{fig:eval-cr}
\end{minipage} \\[-10pt]
\end{figure*} 

\stitle{Live migration of a GPU process (checkpoint-restore). \,}
{\sys} implemented pre-copy-style live migration~\cite{DBLP:conf/nsdi/ClarkFHHJLPW05}:
It first checkpoints the process
via our recopy protocol, then restores it on the target node.
The recopy protocol is necessary because the destination should resume exactly
at the last execution state at the source node.
To avoid redundant data copy from first copying the data from source GPU to the checkpoint,
then from the checkpoint to the target GPU,
we use GPU-direct RDMA~\cite{gpu-direct-rdma} to directly copy the data from the source GPUs' buffers
to the target GPUs' buffers.

\stitle{Fast GPU serverless function startups (restore-mostly). \,}
In serverless computing,
processes are created on demand for each request~\cite{DBLP:journals/corr/abs-1902-03383,ali-serverless-gpu}.
Since each process has an entry function to react to a request,
we take a checkpoint right before the entry, 
so once a request comes, we can restore the process from the checkpoint
using our concurrent restore protocol described in \textsection{\ref{sec:design-restore}}
to avoid the cold start costs~\cite{DBLP:conf/osdi/WeiLW0Y0023,DBLP:conf/eurosys/AoPV22} before entry.

\section{Evaluation}
\label{sec:eval}

\noindent
We have implemented {\sys} in 54,839 LoC in C++ and Rust, 
excluding the CUDA GPU driver extension framework, communication libraries
and integration with CRIU~\cite{criu}.

\stitle{Testbed. \,}
We conducted our experiments on A800 servers,
each with eight NVIDIA A800 (80\,GB HBM and 400\,Gbps NVLink interconnects) GPUs,
two Intel Xeon Gold 6348 CPUs (total 56\,cores),
and 1\,TB of DRAM.
Servers are connected via an RDMA network with 100\,Gbps bandwidth between each GPU
and install CUDA 11.3~\cite{cuda123}.

\stitle{Baselines. \,}
We compare {\sys} with NVIDIA's official
OS-level checkpoint and restore tool---cuda-checkpoint~\cite{nvidia-new-ckpt-driver-repo},
which adopts a stop-the-world approach.
Nevertheless, we found it is extremely slow,
e.g., it cannot fully utilize the PCIe bandwidth to copy data,
see {\fig{fig:eval-cr}}.
Unfortunately, we don't have its source code to analyze.
To compare with the best performance of the stop-the-world C/R,
we implemented Singularity~\cite{DBLP:journals/corr/abs-2202-07848}---the state-of-the-art GPU C/R system,
in our codebase as the baseline.
We have carefully tuned the implementation, 
e.g., we leverage pinned memory to achieve maximum data copy performance.
As shown in {\fig{fig:eval-cr}}, 
our implementation of Singularity has orders of magnitude smaller application stall time than cuda-checkpoint. 
Thus, we will omit a detailed comparison with cuda-checkpoint in later analysis.
Finally,
a recent work CRIUgpu~\cite{DBLP:journals/corr/abs-2502-16631}
builds on cuda-checkpoint to support a unified stop-the-world checkpoint and restore.
Its objective is for functionalities, not performance.
Hence, we omit a comparison because it has the same performance as cuda-checkpoint.

\begin{figure*}[!t]
    \centering
    \includegraphics[left, scale=1.2]{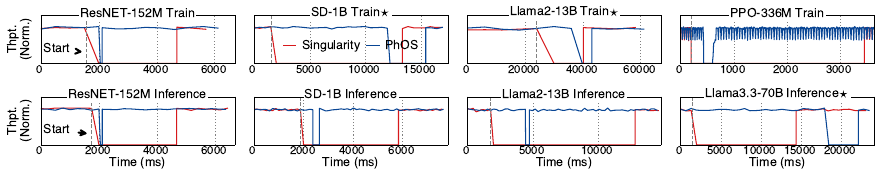} \\[-2pt]
    \begin{minipage}{1\linewidth}
        \caption{\small{%
        The comparison of process migration downtime between machines. 
		($^{\star}$) indicates a multi-GPU setup.%
        }}
    \label{fig:migrate-overall}
\end{minipage} \\[-5pt]
\end{figure*} 

\stitle{Evaluated applications.\,}
We focus on evaluating AI applications---the dominant GPU applications 
to show the effectiveness of {\sys}. 
Table~\ref{tab:app-data} summarizes the evaluated applications. 
We choose workloads of both training and inference,
whose domains span vision tasks (ResNET-152M), 
large foundational models (Llama2-13B and Llama3.3-70B), generation tasks (Stable-Diffusion-1B)
and reinforcement learning (PPO-336M).
These tasks may either use one or multiple GPUs.
We follow common setups for running these applications,
and leave the detailed setup descriptions in the appendix (\textsection{\ref{sec:appendix:setup}}).

\subsection{End-to-end application performance}
\label{sec:eval-e2e}


\nospacestitle{Fault tolerance: evaluated metrics and results. \,}
To evaluate the effectiveness of {\sys} in fault tolerance scenarios,
we choose training applications. 
Following existing works~\cite{DBLP:conf/sosp/WangJZZFNW23,DBLP:conf/eurosys/GuptaKKVGKRS24},
we compare different systems with the following metrics:
checkpoint overhead and wasted GPU time during training.
The checkpoint overhead measures the stall time caused by
checkpointing, while the wasted GPU time measures
the end-to-end training GPU time wasted when compared to a
non-faulty execution.
This includes both the ratio of checkpoint overhead within the overall execution 
as well as the recomputation time
caused by recovering from a stale checkpoint. 
For all systems, 
the checkpoint is stored in host memory to avoid slow storage~\cite{DBLP:conf/sosp/WangJZZFNW23}.

\begin{figure}[!t]
        \centering
        \includegraphics[left, scale=1.0]{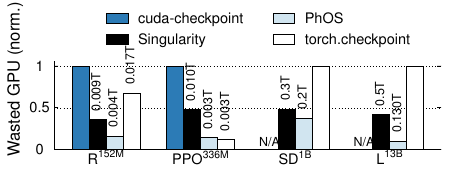} 
        \begin{minipage}{1\linewidth}
        \caption{\small{%
        The wasted GPU time (proportional to the total training time $T$)
        using different
        checkpoint methods for fault tolerance for training workloads.
        To simplify comparison between applications,
        for an application,
        the bar is normalized to the number of systems with the
        maximum wasted time. 
        cuda-checkpoint does not support checkpointing distributed jobs.%
        }}
        \label{fig:eval-app-fault-tolerance}
    \end{minipage} \\[-10pt]
\end{figure}

{\fig{fig:eval-cr}}(a) measures the checkpoint overhead,
with the checkpoints done at the beginning of each training iteration. 
The stall is calculated by first subtracting the total training time with and
without the checkpoint and then normalized to a single iteration time.
We can see that
{\sys} reduced the checkpoint overhead by 70--160\% compared to Singularity
thanks to the concurrent checkpoint.
Notably, on Llama2-13B training,
the overhead of {\sys} is only 185\,ms,
while Singularity is 3.2\,s, bottlenecked by
transferring the GPU checkpoint data (72\,GB) to the host memory (via 32\,GBps PCIe).
For reference, the iteration time is 6.9\,s.

{{\fig{fig:eval-app-fault-tolerance}}} further compares the average wasted GPU time
caused by different checkpoint methods. 
Since the wasted time is related to the failure frequency and the checkpoint frequency, 
we set a failure ratio of one GPU failure per hour reported by various industry reports~\cite{DBLP:journals/corr/abs-2205-01068,DBLP:conf/fast/MohanPC21},
and then calculate the optimal checkpoint frequency for each system
according to the formula in \textsection{\ref{sec:appendix:freq}}.
Note that different systems require distinct checkpoint frequencies:
{\sys} achieves optimal performance with 279 checkpoints per hour,
whereas Singularity performs best at 67.
{\sys} can set a higher frequency thanks to the reduced checkpoint overhead.
The results show that {\sys} saves up to 22--86\% GPU time compared with other approaches:
the higher checkpoint frequency of {\sys}
saves wasted GPU time due to recomputation,
while our concurrent checkpoint minimizes wasted GPU time caused by high checkpoint overhead with high checkpoint frequency.
Interestingly, a well-implemented OS-level GPU C/R even outperforms a simple
user-level checkpointing (torch.checkpoint) that uses PyTorch's \texttt{save}.

\stitle{Live migration: evaluated metric and results. \,}
To evaluate the effectiveness of {\sys} in live migration scenarios,
we measure the downtime when migrating applications between nodes with different systems.
{\fig{fig:migrate-overall}} shows the timeline of application performance during migration.
We can see that {\sys} incurs minimal downtime thanks to the concurrent
execution capabilities.
Notably, {\sys} only incurs 3.3 and 3.7 seconds of downtime for migrating
Llama2-13B training and Llama3.3-70B inference, respectively.
In comparison, Singularity incurs 10.2 and 12.35 seconds of downtime,
which is dominated by copying data from the source GPU to the target.
In Llama2-13B training, copying data through 100\,Gbps RDMA takes at least 9.8 seconds.

\begin{figure}[!t]
        \centering
        \includegraphics[left, scale=1.0]{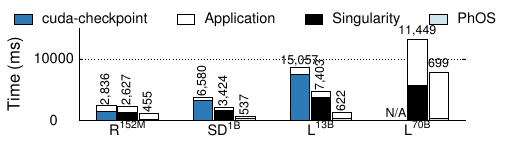} 
        \begin{minipage}{1\linewidth}
        \caption{\small{%
            The breakdown of serverless function execution time in cold starts.     
            Note that all workloads are inference.%
        }}
        \label{fig:eval-app-startup}
    \end{minipage} \\[-10pt]
\end{figure}

\stitle{Serverless function startup: evaluated metric and results. \,}
We evaluate the effectiveness of {\sys} in GPU serverless workloads
by measuring the end-to-end execution time,
i.e., considering both startup and application function execution time~\cite{DBLP:conf/eurosys/AoPV22,DBLP:conf/asplos/DuYXZYQWC20}.
We stored the function checkpoint in host DRAM
and measured the execution time under cold start with OS-level restore.
We choose inference workloads because training workloads are stable and do not
need serverless-style startup.
{\fig{fig:eval-app-startup}} shows the results:
{\sys} has the fastest execution time: on average,
it improves upon cuda-checkpoint and Singularity by 24$\times$ and 16$\times$ respectively,
thanks to the eliminated GPU context creation time, as well as
its ability to hide 
data stalls due to copying. 

\subsection{Performance breakdown and ablation study}
\label{sec:eval-factor}

\begin{figure}[!t]
        \hspace{-1mm}
        \centering
        \includegraphics[left, scale=1.0]{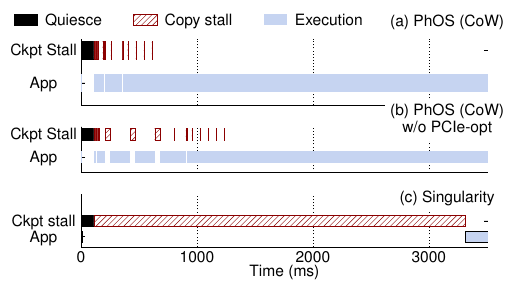} 
        \begin{minipage}{1\linewidth}
        \caption{\small{
            The breakdown of the checkpoint overhead 
            on (a) {\sys} CoW,
            (b) without prioritized PCIe transfer optimization, and 
            (c) Singularity.
            Workload: Llama2-13B Training$^{\star}$.                  
            Note that we omitted the concurrent data copy of {\sys}. 
        }}
        \label{fig:eval-cow-breakdown}
    \end{minipage} \\[-10pt]
\end{figure}

\nospacestitle{Breakdown of the copy-on-write protocol. \,}
{\fig{fig:eval-cow-breakdown}} shows the breakdown of the checkpoint stall caused by different checkpoint methods
on a Llama2-13B training workload.
Other workloads share a similar trend.
All the system checkpoints at the beginning of the iteration: the optimal checkpoint timing,
as analyzed later in \textsection{\ref{sec:eval-timing}}.
We can see that first,
though the quiescing phase will stop the application execution,
its absolute time (100\,ms) is much smaller so this phase has a negligible overhead.
The absolute time is small
because (1) ongoing kernels are short-lived
and (2) coordinating between threads with RDMA to reach a global quiesce
is extremely efficient.
Second, though {\sys} may stall application due to copy-on-write,
the aggregated stalls are small. 

\stitle{Effectiveness of the prioritized application PCIe transfer. \,}
As shown in {\fig{fig:eval-cow-breakdown}},
without our prioritized application PCIe transfer optimization 
described in \textsection{\ref{sec:checkpoint-fast}},
applications suffer from stalls due to PCIe bandwidth competition.
This is because GPU kernels are waiting to load the training set from the CPU,
which is blocked by the checkpoint data copy due to the limited number of GPU DMA engines.

\begin{figure}[!t]
        \hspace{-1mm}
        \centering
        \includegraphics[left, scale=1.0]{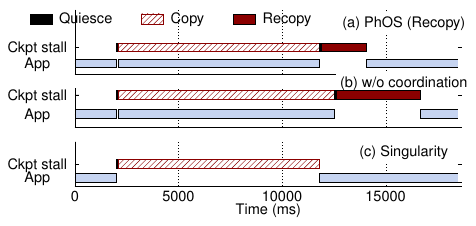} \\[0pt]
        \begin{minipage}{1\linewidth}
        \caption{\small{%
            The breakdown of the checkpoint overhead 
            on (a) {\sys} recopy,
            (b) without coordinated CPU and GPU checkpoint, and 
            (c) Singularity.
            Workload: Llama3.3-70B Inference$^{\star}$.                        
            Note that we omitted the concurrent data copy of {\sys}.%
        }}
        \label{fig:eval-recopy-breakdown}
    \end{minipage} \\[-10pt]
\end{figure}

\stitle{Breakdown of the recopy protocol. \,}
{\fig{fig:eval-recopy-breakdown}} shows the breakdown of executing our recopy protocol
on a Llama3.3-70B inference workload$^{\star}$.                 
We can see that the downtime of {\sys} is dominated by the recopy time.
Compared to CoW, the recopy downtime is longer because it 
not only ensures that the checkpoint is correct,
but it is also fresh.
Nevertheless, the downtime is still orders of magnitude smaller than a stop-the-world
approach (Singularity) because the recopied data is much smaller (2.1 vs. 9.7 seconds).

\stitle{Effectiveness of the coordinated CPU and GPU checkpoint. \,}
{\fig{fig:eval-recopy-breakdown}} further evaluates the effectiveness of our coordinated CPU and GPU checkpoint optimization
in \textsection{\ref{sec:checkpoint-fast}}, aiming to reduce the amount of dirty data for the recopy phase.
Thanks to the optimized GPU concurrent copy timing,
{\sys} enjoys 47\% smaller recopy time than without optimization (b).
The recopied data reduces from 50 to 27\,GB per GPU.
Since the copy is a bulk load procedure, the reduced transferred size
directly translates to a reduced recopy time.

\begin{figure}[!t]
        \hspace{-1mm}
        \centering
        \includegraphics[left, scale=1.0]{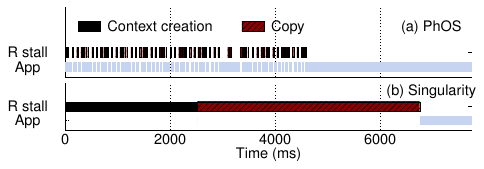} 
        \begin{minipage}{1\linewidth}
        \caption{\small{%
            The breakdown of the restore 
            on (a) {\sys} concurrent restore,
            and (b) Singularity. 
            Workload: Llama2-13B inference.      
            Note that we omitted the concurrent data copy of {\sys}. 
            R stall stands for \uline{r}estore stall time.%
        }}
        \label{fig:eval-restore-breakdown}
    \end{minipage} \\[-15pt]
\end{figure}

\stitle{Breakdown of the concurrent restore. \,}
{\fig{fig:eval-restore-breakdown}} shows the breakdown of concurrent restoring a
Llama2-13B inference process using {\sys}.
We can see that compared to the stop-the-world restore (b),
the improvements of {\sys} come from two factors:
(1) the eliminated context creation
and (2) overlapping the data copy with the kernel execution.
Note that during application execution,
{\sys} still copies the data in the background:
this perfectly aligns with how inference processes access data nowadays:
before executing the first layer of inference, we can concurrently restore the
second layer's data.
Unlike previous work~\cite{10.1145/3698038.3698510,pipeswitch}, {\sys} achieves so transparently. 

\begin{figure*}[!t]
    \centering
    \includegraphics[left, scale=1.25]{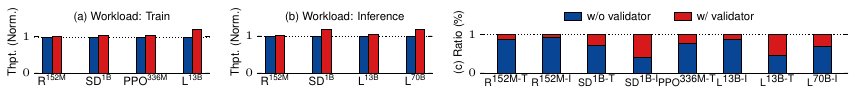} 
    \begin{minipage}{1\linewidth}
            \caption{\small{%
                The analysis of runtime validator overhead on
                (a) training and (b) inference workloads, along with 
                (c) the ratios of kernels instrumented with the validator.
                Training is abbreviated as \textbf{-T} and inference as \textbf{-I}.%
        }}
    \label{fig:eval-instrument}
\end{minipage} \\[-5pt]
\end{figure*}

\stitle{The runtime overhead of tracing GPU read and write set. \,}
{\fig{fig:eval-instrument}}(a) and (b) report the runtime overhead
of the runtime validator on different applications.
We observe a relatively small slowdown of 1--12\% for various workloads.
The observable overhead is small because:
(1) the additional check only happens when the kernel accesses the global memory,
which is less frequent for high computational efficiency;
and (2) the instrumented kernels are only a small portion of the
overall kernels in the workloads.
{\fig{fig:eval-instrument}}(c)
reports the number of instrumented kernels invoked during the concurrent checkpoint and restore:
We see that most of the kernels are not instrumented,
e.g., in Llama2-13B inference, only 12\% of the invoked kernels are instrumented.

\subsection{Impact of on-device memory used on performance}
\label{sec:eval-mem-capacity}

\begin{figure}[!t]
        \centering
        \includegraphics[center, scale=1.05]{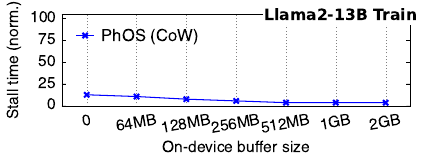} 
        \begin{minipage}{1\linewidth}
        \caption{\small{%
            The analysis of the copy-on-write overhead with different 
            on-device buffer sizes.%
        }}
        \label{fig:eval-ondevice-buffer-size}
    \end{minipage} \\[-10pt]
\end{figure}

\noindent
{\fig{fig:eval-ondevice-buffer-size}} shows the impact of the on-device buffer size
on the checkpoint overhead when checkpointing a Llama2-13B training workload.
We choose this workload because it has the largest memory footprint
so little room is left for the on-device buffer (see Table~\ref{tab:app-data-full}).
We report the increased stall time caused by the checkpoint,
where the time is normalized to the stall time with our maximum
configurable on-device buffer size (2\,GB).
We see that checkpoint overhead generally decreases with the on-device buffer size,
but the difference is small,
e.g., even without on-device buffer,
{\sys} incurs up to 9.1\,\% more stalls.
This is because not all copy-on-write operations require stalling the kernel,
so the slow copy through PCIe with insufficient on-device buffer
can be hidden.

\subsection{Impact of checkpoint timing on performance}
\label{sec:eval-timing}

\begin{figure}[!t]
        \centering
        \includegraphics[left, scale=1.0]{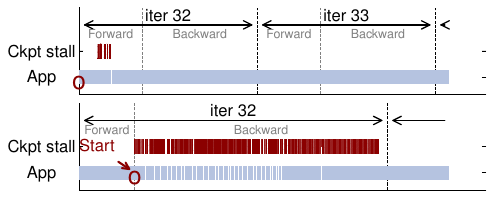} \\[5pt]
        \begin{minipage}{1\linewidth}
        \caption{\small{
            Impact of checkpoint timing on the performance of Llama2-13B training.
            The blank in App is the stall caused by the CoW. 
            The red indicates the data copy time. 
            Note that we omitted the concurrent data copy of {\sys}. 
        }}
        \label{fig:eval-ckpt-timing}
    \end{minipage} \\[-15pt]
\end{figure}

\noindent
{\fig{fig:eval-ckpt-timing}} analyzes how the checkpoint timing
affects the performance of {\sys} on the CoW protocol.
We omit analysis of the recopy protocol as it is similar.
We control the timing with our SDK (see \textsection{\ref{sec:appendix:sdk}}).
We run Llama2-13B training workload and choose two times: 
(1) at the beginning of an iteration (before the forward pass),
(2) at the end of the iteration (at the update phase).
We measure the performance specifically on the 32$^{th}$ iteration to fully
warm up the application.
We can see that checkpoint at (1) is more efficient because
few buffers are updated at the beginning: only activation buffers are updated (see also \textsection{\ref{sec:trace-granularity}}).
Thus, if we can finish the checkpoint before reaching the update pass,
we meet a few stalls due to CoW.
In (1), 
copying 2.3\,GB of data via 32\,GBps PCIe\footnote{\footnotesize{The measured
bandwidth is 25\,GBps, slightly below the hardware limit.}}
takes 185\,ms, while the iteration time is 6.3\,s.

\subsection{A close look at GPU read and write traced by {\sys}}
\label{sec:trace-granularity}

\begin{figure}[!t]
        \centering
        \includegraphics[left, scale=1.0]{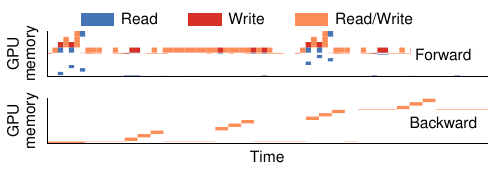} 
        \begin{minipage}{1\linewidth}
        \caption{\small{
           The heatmap of the read and write sets traced by our validation method. 
		   Workload: Llama2-13B training$^{\star}$.
        }}
        \label{fig:eval-heatmap}
    \end{minipage} \\[-10pt]
\end{figure}

\noindent
{\fig{fig:eval-heatmap}} reports the GPU accesses traced by {\sys} 
on Llama2-13B training. 
First, we can see that though {\sys} traces accesses at the buffer level,
it is still fine-grained enough to support an efficient
CoW and recopy, because
GPU applications typically allocate and write buffers in a fine-grained way.
Moreover, the access patterns differ across different phases of execution, 
so the checkpoint timing is important.

\begin{table}[!t]
    \vspace{1mm}
    \begin{minipage}{1\linewidth}
        \caption{\small{%
        The success rate of our speculation-based
        GPU read and write sets tracing beyond the evaluated applications.%
        }}
    \label{tab:eval-validate}
    \end{minipage} 
    \begin{minipage}{1.\linewidth}
    \centering
    \small{
    \ra{1.1}
    \begin{tabular}{l|r@{~}l|r@{~}l}
        \toprule
        \textbf{GPU Apps} & \textbf{\#Kernels /} & \textbf{\#Failures} & \textbf{\#Instances /} & \textbf{\#Failures} \\
        \midrule
        Rodinia~\cite{DBLP:conf/iiswc/CheBMTSLS09}  &  44 / & {\color{red}1}  &  48,610 / & {\color{red}20} \\
        Parboil~\cite{stratton2012parboil}          &  18 / & {\color{blue}0} &  43,473 / & {\color{blue}0} \\
        vLLM~\cite{vllm}                            &  66 / & {\color{blue}0} &  13,625 / & {\color{blue}0} \\
        TVM~\cite{DBLP:conf/osdi/ChenMJZYSCWHCGK18} & 607 / & {\color{blue}0} & 186,244 / & {\color{blue}0} \\ 
        FlashInfer~\cite{flashinfer}                &  69 / & {\color{blue}0} &  15,265 / & {\color{blue}0} \\
        \bottomrule        
    \end{tabular} 
    } 
\end{minipage} \\[-10pt]
\end{table}  

\subsection{A study of the feasibility of speculation}
\label{sec:eval-speculation}

\noindent
For all the common GPU applications evaluated by {\sys},
we don't meet any speculation failure. 
To evaluate the feasibility of our approach,
we conduct a comprehensive study across diverse GPU applications
including supercomputing workloads (Rodinia~\cite{DBLP:conf/iiswc/CheBMTSLS09} and Parboil~\cite{stratton2012parboil}), 
dynamically generated kernels from AI compilers (TVM~\cite{DBLP:conf/osdi/ChenMJZYSCWHCGK18}), 
and highly optimized handwritten kernels (vLLM~\cite{vllm} and FlashInfer~\cite{flashinfer}) in Table~\ref{tab:eval-validate}. 
To evaluate the success rate of our speculation, 
we run typical tasks provided by them (we term each run of a task \emph{instance}). 
For Rodinia and Parboil, 
we ran all benchmarks except for ones that had bugs or used outdated CUDA APIs (e.g., dwt2d).
For TVM, we ran inference on their available models other than our previous evaluated ones (e.g., DenseNet, YOLOv4).
Out of all instances, only one kernel (from Rodinia) failed our speculation---a fairly dated supercomputing application. 
The failure occurred because a kernel reads a buffer referenced by a global variable not listed in the launch arguments.

\section{Discussion}
\label{sec:dislim}

\nospacestitle{Supporting advanced GPU features. \,} 
CUDA graph~\cite{cuda-graph} is an advanced GPU feature 
that allows the CPU to submit a batch of kernels. 
{\sys} supports tracing reads and writes of kernels within CUDA graph,
inspired by Medusa~\cite{DBLP:conf/asplos/ZengXGCL25}.
Specifically,
CUDA graphs can be utilized as follows:
(1) explicitly invoking APIs such as \texttt{cudaGraphAddKernelNode},
or (2) implicitly constructing them through the driver using \texttt{cudaStreamBeginCapture}.
Both methods require explicit driver API calls, so they are compatible with our API-hooking-based tracing. 

\stitle{GPU hardware extensions. \,}
{\sys} is the first to show the effectiveness of concurrent GPU checkpoint and restore
as well as how to realize it in major GPUs.
While hardware extensions like GPU dirty bit
could simplify and accelerate {\sys}'s implementation,
relying on hardware extensions is less flexible,
e.g., a hardware dirty bit alone cannot support our other protocols like soft copy-on-write.

\stitle{Application-provided kernel access information. \, }
Providing the kernel's memory access to {\sys} via developer annotation
or just-in-time (JIT) compilation
would fully utilize {\sys}'s capabilities.
Yet modifying numerous existing custom kernels for the annotation, as
well as designing an effective JIT for checkpointing, is non-trivial.
How to develop kernel memory access tracing methods beyond our speculative approach
is left as our future work.

\section{Related work}
\label{sec:related}

\nospacestitle{Checkpoint and restore. \,}
{\sys} advances the research on checkpoint and restore (C/R), specifically focusing
on efficient OS-level GPU C/R~\cite{egwutuoha2013survey,litzkow1997checkpoint,
keykos,eros,Aurora,criu,laadan2010linux,hargrove2006berkeley,DBLP:conf/sosp/Wu0M023,
DBLP:journals/corr/abs-2502-16631},
a feature previously available only on CPUs. 
GPU snapshot~\cite{DBLP:conf/ics/LeeSHTKE19} realizes dirty bits through simulated GPU hardware extensions,
which to the best of our knowledge, no actual hardware implementation currently exists. 
gCROP~\cite{10.1145/3698038.3698510}
uses concurrent restore to accelerate process startup (only) on AMD GPUs, 
but requires significant application modifications to ensures correctness.
Their approach is infeasible on closed-source GPUs (e.g., NVIDIA GPUs),
does not support multi-GPU applications, and most importantly,
lacks support for concurrent GPU checkpoint.
In contrast, our work enables both concurrent GPU checkpoint and restore 
without application modifications through our novel validated speculation approach. 
{\sys} is the first system to support both concurrent OS-level checkpoint and restore 
for GPU applications 
across multiple GPUs and major GPU vendors like NVIDIA GPUs.

\stitle{Analyzing GPU kernels. \,}
Many existing works analyze GPU programs (kernels),
either statically or during runtime~\cite{DBLP:conf/nfm/ChiangGLR13,DBLP:conf/sigsoft/LiG10,DBLP:journals/toplas/BettsCDKQTW15,DBLP:conf/oopsla/BettsCDQT12,DBLP:conf/sosp/KamathB21,DBLP:conf/pldi/LeungGAGJL12,DBLP:conf/ppopp/LiLSGGR12,Mai2023HONEYCOMB}.  
{\sys} takes a different approach: instead of analyzing the program,
we analyze the kernel arguments for speculation. 
This avoids the incompleteness inherent in GPU kernel analysis 
(as GPU kernels are Turing complete)
and works well for concurrent GPU checkpoint and restore. 

\section{Conclusion}
\label{sec:concl}

\noindent
We contribute the first concurrent OS-level GPU checkpointing and restoring system
using validated speculation, and demonstrate 
its effectiveness in various critical downstream applications, 
including fault tolerance, live migration, and fast startup.
Our current prototype is built on NVIDIA GPUs,
but our methodologies are generalizable to other accelerators, 
and we plan to support these devices in the future. 

\section*{Acknowledgment}
\noindent
We sincerely thank our shepherd Ashvin Goel and 
the reviewers from SOSP'24, OSDI'24 and SOSP'25 for
their insightful feedback. 
We are grateful to Zhiyuan Dong, Dingyan Zhang and Xiating Xie for 
their valuable comments on the early draft of this paper. 
We also thank Shaoxun Zeng for sharing details on adapting {\sys} 
to support CUDA graph.
This work was supported in part by
the National Key Research \& Development Program of China (No. 2022YFB4500700),
the National Natural Science Foundation of China (No. 62202291 and 62272291),
as well as a research grant from Huawei.

\balance

\small{
\bibliographystyle{acm}
\bibliography{pos}
}

\clearpage

\nobalance
\appendix
\section{Appendix}
\label{sec:appendix}

\begin{table*}[!t]
    \vspace{1mm}
    \begin{minipage}{1\linewidth}
    \caption{\small{%
        Detailed setups of applications evaluated in \textsection{\ref{sec:eval}}.
        ($^{\star}$) indicates a multi-GPU setup.
        PPO is training-only, and our testbed cannot run L$^{\text{70B}}$ 
		training.%
        }}
    \label{tab:app-data-full}
    \end{minipage} 
    \begin{minipage}{1\linewidth}
    \centering
    \small{
            \ra{1.1}
            \begin{tabular}{@{}l|cc|cc|cc|cc|cc@{}}
                \toprule
                & \multicolumn{2}{c|}{\textbf{ResNET-152M}}
                & \multicolumn{2}{c|}{\textbf{PPO-336M}}
                & \multicolumn{2}{c|}{\textbf{Stable-Diffusion-1B}}
                & \multicolumn{2}{c|}{\textbf{Llama2-13B}} 
                & \multicolumn{2}{c}{\textbf{Llama3.3-70B}} \\ 
                & \multicolumn{2}{c|}{(R$^{\text{152M}}$)}
                & \multicolumn{2}{c|}{(PPO$^{\text{336M}}$)}
                & \multicolumn{2}{c|}{(SD$^{\text{1B}}$)}
                & \multicolumn{2}{c|}{(L$^{\text{13B}}$)} 
                & \multicolumn{2}{c}{(L$^{\text{70B}}$)} \\ 
                \midrule
                \textbf{Application}
                & \textbf{Training} & \textbf{Inference}               
                & \multicolumn{2}{c|}{\textbf{Training}}               
                & \textbf{Training}$^{\star}$ & \textbf{Inference}     
                & \textbf{Training}$^{\star}$ & \textbf{Inference}     
                & \textbf{} & \textbf{Inference}$^{\star}$ \\  
                \textbf{Library}
                & \multicolumn{2}{c|}{Torchvision~\cite{torchvision} } 
                & \multicolumn{2}{c|}{OpenAI Gym~\cite{gym}}            
                & \multicolumn{2}{c|}{HuggingFace~\cite{huggingface}}   
                & \multicolumn{2}{c|}{Meta Llama 2~\cite{llama}}        
                & \multicolumn{2}{c}{Meta Llama 3.3~\cite{llama3}} \\     
                \midrule
                \textbf{\#GPUs}
                & 1 & 1     
                & \multicolumn{2}{c|}{1}           
                & 8 & 1             
                & 8 & 1             
                & \multicolumn{2}{c}{8} \\        
                \makecell[l]{\textbf{Memory usage} \\(per GPU)}
                & 1.8\,GB & 1.7\,GB   
                & \multicolumn{2}{c|}{5.9\,GB}        
                & 70.6\,GB & 8.9\,GB     
                & 73.6\,GB & 55.4\,GB 
                & \multicolumn{2}{c}{70.8\,GB} \\    
                \makecell[l]{\textbf{\#Buffers} \\(per GPU)}
                & 209 & 195             
                & \multicolumn{2}{c|}{75}              
                & 445 & 234         
                & 413 & 347             
                & \multicolumn{2}{c}{718}     \\      
                \makecell[l]{\textbf{\#Kernels} \\(active)}
                & 13 & 8                
                & \multicolumn{2}{c|}{41}              
                & 51 & 50               
                & 36 & 74               
                & \multicolumn{2}{c}{73} \\           
                \bottomrule
            \end{tabular}
    }
    \end{minipage} 
\end{table*}

\subsection{Determine the optimal checkpoint frequency}
\label{sec:appendix:freq}

\nospacestitle{Problem formulation. \,}
Consider a distributed computing job using $N$ GPUs, where
each GPU fails $F$ times per hour.
Assume these GPU failures are independent and identically distributed, 
following a uniform distribution over the entire computation interval ($T$). 
with {\sys}, the checkpoint overhead is $O$. 
When a GPU fails, {\sys} stops all GPU processes
and restores from the latest checkpoint, with a restore time of $R$ for all GPUs. 

The total GPU hours wasted due to checkpoint overhead is: 

\[
N\,O\,f\,T \,.
\]

\vspace{2mm}
The total GPU time wasted due to failures at a checkpoint frequency of $f$ times per hour is: 

\[
\Bigl(N\,F\,T\Bigr)
\;\times\;
\Bigl(R\;+\;\tfrac{N}{2f}\Bigr) \,.
\]

\vspace{2mm}
\stitle{Put it all together. \,} 
The total GPU hours wasted due to checkpoint overhead and failure recovery is: 

\[
\Bigl(N\,F\,T\Bigr)
\;\times\;
\Bigl(R\;+\;\tfrac{N}{2f}\Bigr)
\;+\;
N\,O\,f\,T \,.
\]

\vspace{2mm}
\stitle{Solving the optimal $f$. \,}
For a given job, the frequency of failures ($F$), 
the checkpoint overhead ($O$), and the restore time ($R$) are static 
and can be profiled online~\cite{DBLP:conf/fast/MohanPC21}.
Therefore, we only need to determine the optimal checkpoint frequency $f$ 
to minimize wasted GPU hours.
By differentiating the total GPU hours wasted with respect to $f$, we derive: 

\[
f^*
\;=\;
\sqrt{\frac{N\,F}{2\,O}} \,.
\]

\vspace{2mm}
\noindent {\sys} sets the checkpoint frequency to $f^*$ for each computing job.

\vspace{1mm}
\subsection{{\sys} software development kit (SDK)}
\label{sec:appendix:sdk}

\definecolor{myred}{HTML}{C00000}

\begin{figure}[!t]
        \vspace{1mm}
        \begin{minipage}{1\linewidth}
        \centering    
        \includegraphics[width=1.01\linewidth, trim=0.25cm 10.4cm 21cm 0.25cm, clip]{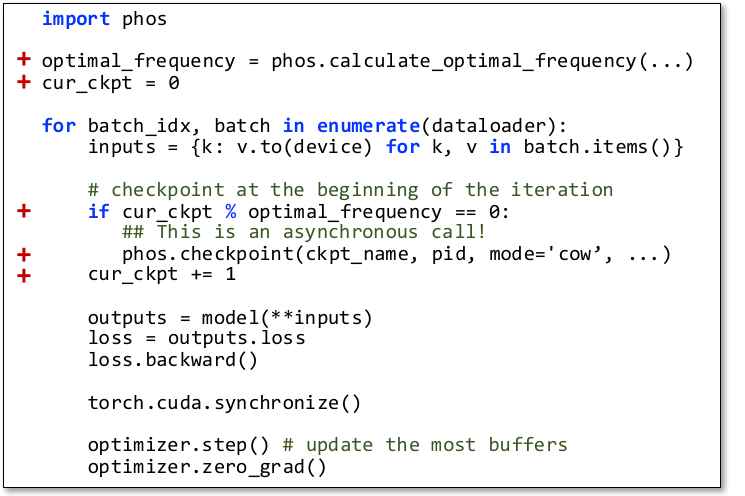} \\[2pt]
        \end{minipage} \\[5pt]
        \begin{minipage}{1\linewidth}
        \caption{\small{%
        An illustration of how applications can use the {\sys} SDK
        to control checkpoint timing.
        Lines marked with ``\textcolor{myred}{+}'' indicate code that interacts with {\sys}.
        }}
        \label{fig:sdk}
        \end{minipage} 
        \end{figure} 

\noindent
Applications can explicitly call our SDK to control checkpoint timing 
and the protocol used, as shown in {\fig{fig:sdk}}.
Our OS-level design merely requires minimal code changes---just one line of code.
For example, the code instructs {\sys} to checkpoint the process
at the beginning of a training iteration.
Since the \texttt{checkpoint} is executed asynchronously, 
it will not block application execution unless the last checkpoint is still in progress.

\vspace{1mm}
\subsection{Detailed setups of our evaluated applications}
\label{sec:appendix:setup}

\noindent
Table~\ref{tab:app-data-full} shows the detailed buffer allocation information
as well as the number of GPU kernels information of our evaluated applications.
For training workloads: 
ResNET uses the CIFAR-10 dataset with a batch size of 32;
PPO uses the gym code~\cite{gym}; 
Stable Diffusion uses the SD v1-4 model with a batch size of 1,536 per GPU;
and Llama uses a distributed configuration of 8TP (Tensor Parallelism) and 1DP (Data Parallelism)
with a batch size of 4 due to GPU resource constraints.
All training workloads use the AdamW optimizer.
We use the same training datasets to evaluate the inference workloads.

\end{document}